\documentclass[10pt,letterpaper]{article}
\usepackage{geometry}
\usepackage[utf8]{inputenc}
\usepackage{amsmath}
\usepackage{amssymb}
\usepackage{aastex}
\usepackage[round, comma, sort&compress]{natbib}
\usepackage{nameref,hyperref}
\usepackage{graphicx}
\usepackage{lastpage,fancyhdr}
\usepackage{epstopdf}
\usepackage{float}
\usepackage{enumitem}
\usepackage{changepage}
\usepackage[aboveskip=1pt,labelfont=bf,labelsep=period]{caption}

\usepackage{microtype}
\DisableLigatures[f]{encoding = *, family = * }

\usepackage{color}
\definecolor{Gray}{gray}{.25}

\usepackage[explicit]{titlesec}
\titleformat{\section}
{\normalfont\normalsize\bfseries\centering}{\thesection}{1em}{\MakeUppercase{#1}}

\newlist{arrowlist}{itemize}{1}
\setlist[arrowlist]{label=$\rightarrow$}

\begin{document}
\vspace*{0.35in}

\begin{center}
	\begin{center}
{\Large
\textbf\newline{Magnetar Signature - The U Curve}
}\\
\vspace{4mm}
Vikram Soni\textsuperscript{1,2,*},
Dipankar Bhattacharya\textsuperscript{2},
Sameer Patel\textsuperscript{3},
Sajal Gupta\textsuperscript{4},
Prasanta Bera\textsuperscript{2}
\end{center}
\bigskip
\bf{1} Centre for Theoretical Physics, Jamia Millia Islamia, New Delhi 110025, India
\\
\bf{2} Inter University Centre for Astronomy and Astrophysics, Post Bag 4, Pune
411007, India
\\
\bf{3} Department of Physical Sciences, P D Patel Institute of Applied Sciences, Charotar University of Science \& Technology, Changa 388421, Gujarat, India
\\
\bf{4} Department of Physical Sciences, Indian Institute of Science Education and Research, Kolkata, Mohanpur 741246, India
\\
\bigskip
* vsoni.physics@gmail.com

\end{center}

\section*{Abstract}
This work looks at some definitive signatures of magnetars, in particular of period closures accompanied by a decline of X-ray radiation in two models. We review some of the previous works which are based on the well known dynamo model in which the star is born with a period of a few milliseconds at high temperatures. In such a convection regime the dynamo mechanism can amplify the the magnetic fields to the magnetar value. This is in contrast to a screened core model which posits that a high density phase transition occurs in the inner core of magnetars that dynamically aligns all the neutron magnetic moments producing a large magnetic field in the core. The accompanying change of flux gives rise to shielding or screening currents in the surrounding high conductivity plasma that do not permit the field to exit to the surface. Ambipolar diffusion then transports the field to the crust dissipating energy in neutrinos and X-rays. The up-welling  field cleaves the crust resulting in flares and X-ray radiation from ohmic dissipation in the crust till the screening currents are spent and the surface polar field attains its final value. In the dynamo model the polar magnetic field decreases with time whereas in our screened model it increases to its final value. One consequence of this is that in the latter model, as a function of time and period, the ratio of the dipole radiation loss, $\dot E$ to the X-ray luminosity, $ L_X$, is a `U' curve, indicating that it is the exponential decline in $L_X$, that brings closure to the periods that are observed for magnetars.

\section{Introduction}
\label{sec1}

Magnetars are usually identified as isolated neutron stars with the largest surface polar magnetic fields ($10^{14} - 10^{15}$ G) and that have spin down ages of $10^{3} - 10^{5}$ years. Over this period, they emit a quiescent radiative luminosity of $10^{34} - 10^{36}$ erg s$^{-1}$. Besides, some of them emit repeated flares or bursts of energy typically of $10^{42} - 10^{44}$ erg, and at times of even higher intensity \citep{Hurley}. The periods of magnetars fall in a surprisingly narrow window of $2-12$ s. \emph{However, there are exceptional magnetars which may not share all these features.}\\

At such large periods, the energy emitted in both quiescent emission and flares far exceeds the loss in their rotational energy through dipole radiation. The most likely energy source for these emissions is their magnetic energy, yet there is no evidence of a decrease in their surface (polar) magnetic fields with time \citep{Thompson}. There have been many attempts to explain some of this physics of which the most popular  is the magnetar model of Duncan and Thompson \citep{DuncanThompson,ThompsonDuncan}, which is known as the {\em dynamo mechanism for magnetars}. This model requires the collapse of a large mass progenitor to a star which starts life with a period close to a few milliseconds. At the high temperatures generated by the collapse process, such a fast rotation can amplify the inherited pulsar valued field of $10^{12}$ G to $10^{15}$ G or more. However, as described below, several observations on magnetars are hard to understand from such a  model.\\

If magnetars are born with their high surface magnetic fields, then, after a flare, one would expect a decrease in the magnetic energy and consequently a {\em decrease} in the polar magnetic field  of the magnetar {accompanied by }a fall in the dipole radiation mediated spin down rate. However, sometimes the opposite is seen: the spin down rate {\em increases} after the flare indicating an increase in the surface magnetic field \citep{DarDR,Dar,Marsden}. Further, inspite of the magnetic energy loss  from steady X-ray emission, the magnetic field of magnetars appears to remain high all the way till the end of spin down, when the stars have the largest periods.\\

Another enigmatic feature of magnetars is that no magnetar with a period larger than 12 seconds has been observed -  inspite of high magnetic fields at such large  periods (of the order of ten seconds). For accreting binary X-ray pulsars with smaller magnetic fields, it is common to find periods as large as even a minute. This has motivated dynamical models \citep{Rea1} that will dissipate crustal  magnetic fields through a resistive layer and simultaneously rapidly diminish $ \dot P$  leading to a limiting value of P (of the order of $10-20$ s).\\

We note that at the high polar crustal and surface magnetic fields ($B_{polar} > 10^{13 -14}$ G), the Landau radius becomes smaller than the Bohr radius. As the crustal magnetic field increases further, the atoms will deform  to cigar like atoms (ions)  with their long axis aligned along the magnetic field. Ions that are strung along  the crystal axis along the magnetic field can accumulate electrons in Landau levels in the transverse direction. At densities where  the Landau radius is appreciably less than the inter ionic distance, the Coulomb interaction will bind the electrons to the ions to make cylindrical atoms which are neutral. In this regime, the Coulomb interaction in the transverse (to the field) direction becomes weak. The overlap between the electron wave functions between sites in the transverse direction diminishes making the crystal  weak in the direction transverse to the field. The screening length also becomes smaller than the inter atomic distance \citep{ShaRed,Bed}. Thus the transverse conductivity also decreases. This is indicated in detailed calculations \citep{Israeli, ourpreprint} which show that the transverse conductivity goes inversely as the square of the polar magnetic field for high polar fields, $B_{polar} > 10^{13}$ G. The transverse crustal currents can then dissipate more effectively via Ohm's law.  As the density gets higher and the inter atomic/ionic distance becomes smaller, we cross over from a regime of neutral atoms to an ionic regime restoring some conductivity.\\

In an earlier work \citep{DipSoni,mag}, it was shown that it may be  possible to explain many unusual features of magnetars if they have a core with a large magnetic moment density, created  at birth. Initially the core magnetic field is shielded by the electron plasma in and around the core. In time, the shielding currents dissipate till finally the field emerges at the surface of the star.\\

In this work, we try and compare some distinguishing properties of two models for magnetars:


\begin{enumerate}
	
	\item In the dynamo model \citep{DuncanThompson,ThompsonDuncan}, at birth, we have a fast rotating star with a period of a few milliseconds and high temperatures. In such a  convection regime the dynamo mechanism can  amplify the the magnetic fields to the high value. In this model  the magnetar is born with the high polar/poloidal magnetic field,$B_{polar}$, at the surface - this is in contrast to the model 2 below.
	
	\item The screened core model \citep{DipSoni,mag}, which posits that a high density phase transition occurs in the inner core of magnetars that aligns all the neutron magnetic moments producing a high magnetic field in the core. This field results in a change of flux in the $n$, $p$, $e$ plasma in the outer core of very high conductivity, where shielding or screening currents are set up that do not permit the field to exit to the surface. Ambipolar diffusion then transports the field to the crust simultaneously dissipating energy in neutrinos and X-rays. As the screening currents dissipate and the core field  reaches the inner crust, the field gradient between the inner crust and the surface cleaves the crust resulting in flares and letting the field into the crust. As pointed out in the previous paragraph, when the field in the crust increases beyond $B_{polar} > 10^{13-14}$ G, the Landau radius becomes smaller than the Bohr radius and a change to cigar like  ions occurs. Ions that are strung along  the crystal axis along the magnetic field can attract electrons and accumulate Landau levels in the transverse direction. This decreases the conductivities in the directions that are transverse to the polar magnetic field thereby increasing the ohmic dissipation and the radiative X-ray luminosity.
	
\end{enumerate}

Whereas it is the dynamo currents  that dissipate in the model $1$, in the screened core model ($4$), it is the screening currents that dissipate. The effect on the polar magnetic field is the opposite. In the dynamo model, $B_{polar} $ goes down through dissipation with time. In the screened core model, $B_{polar} $ goes up as the screening currents dissipate. This is one feature that is ubiquitous is that $B_{polar}$ increases with time for the screened core model and decreases with time for the dynamo model.\\


The plan of this paper is as follows. In section \ref{sec2}, we review some models of dissipation of the crustal currents and the polar magnetic field. We compute the consequent X-ray luminosities and compare them with the observed values for magnetars. In section \ref{sec3}, we examine the dependence of conductivity tensor and ohmic dissipation time scale for high magnetic fields. In sections \ref{sec4.1} and \ref{sec4.2}, we look at the X-ray and dipolar radiation luminosities, periods and decay timescales in the dynamo model and the screened core model. We find that in the screened core model, the ratio of the dipole radiation loss $\dot E$ to the X-ray luminosity ($L_X$) is a `U' curve as a function of increasing spin down age and period. Finally, in section \ref{sec6}, we try and understand the observed cut-off in the period of magnetars and end with a discussion.

\section{Crustal current and polar magnetic field dissipation and X-ray luminosity $\mathbf{L_X}$ for some models}
\label{sec2}

\subsection{Estimation of crustal magnetic field energy decay rate from the  magnetar spin down age}

Let us first make a simple estimate of the rate of magnetic energy dissipation in the crust. We assume the crust to be a shell of thickness, $\Delta R_{cr}$, in which the magnetic fields diminish with time as the crustal currents dissipate. We shall give some estimates which, though qualitative, capture some of the physics.\\

The amount of magnetic energy stored in the crust is given roughly by the expression

\begin{equation}
E_{stored} \sim \frac{B^2}{8\pi}\cdot 4\pi R^2\cdot \Delta R_{cr}~\text{,}
\end{equation}

where $B$ is the observed surface polar field, $R$ ($10$ km) is the radius of the star and $\Delta R_{cr}$ is the thickness
of the crust, which we take to be of the order of $1$ km. As the field drops to a fraction of its original value, this energy  will be dissipated and released as X-ray luminosity.\\

We assume that the  typical time-scale in which this happens is given by the spin-down age of the magnetars,

\begin{equation}
\tau_{SD} = \frac{P}{2\dot P}
\end{equation}

Thus, a crude estimate for the `average' X-ray luminosity is given by dividing the stored energy magnetic by the spin down age,

\begin{equation}
\label{avg.l_x}
L_X  =  \frac{E_{stored}}{ \tau_{SD}}
\end{equation}

Equation~\ref{avg.l_x} can be found in \citep{mag}. Hence,

\begin{equation}
L_X^{calc} \sim \frac{E_{stored}}{\tau_{SD}} = R^2\Delta R_{cr}\times B^2 \times \frac{{\dot P}}{P}
\end{equation}

Now using the relation below which is based on the dipole formula,

\begin{equation}
\label{dipole}
{\dot P} =\frac{2}{3}\frac{4\pi^2}{P}\frac{B^2R^6}{c^3I}~\text{,}
\end{equation}

we find that for magnetic fields, $B_p \sim 4 \times 10^{14}$ G, the magnetar data indicates average periods of  $\sim 8$ s. On substitution, this yields an average crustal X-ray luminosity

\begin{equation}
L_X^{calc}  \sim \frac{E_{stored}}{\tau_{SD}} \sim 3 \times 10^{34}~\text{erg s$^{-1}$}
\end{equation}

The instantaneous luminosity when the periods reach magnetar ages (periods) will, of course, be less than the average.\\

The dipolar spin down radiation energy loss rate for magnetars with such magnetic fields and periods is then given by

\begin{equation}
\dot E = 4 \pi^2 I \frac{\dot P}{P^3} = I \omega \cdot \dot\omega \sim \frac{ 4\pi^2 I \times\ 10^{-39} \cdot {B(t_H)}^2}{P^4} \sim 0.16 \times 10^{34}~\text{erg s$^{-1}$}
\end{equation}

\subsection{Estimation of Magnetic Field Energy Decay Rate Due to Ohmic Resistance and the Hall Effect \citep{WoodHoll}:}
\label{sec2.2}

The work of \cite{WoodHoll,GourCumm,WoodHollPC} computes the magnetic field evolution in the crust via the Hall effect. They find, for example, that starting with polar fields of  $B_0 \sim 4 \times 10^{14}$ G at the outer crust (and comparatively much smaller fields at the inner crust), reduce to half their value in $0.4$ Myr (million years). The crust thickness is taken to be about $1$ km. The Hall drift term cannot directly give rise to dissipation, but can convert polar fields to toroidal fields and also give rise to currents loops of all sizes. The smaller loops can then undergo faster Ohmic dissipation.\\

From the results of \cite{WoodHoll,WoodHollPC}, we can make a crude estimate of the average rate of energy dissipation at the Hall time, which is given below.\\

The Hall timescale is given by,
\begin{equation}
t_H = \frac{C}{B_0}
\end{equation}

This is evident from \citep[eq. 9]{WoodHoll,GourCumm,WoodHollPC} , where $B_0$ is the initial  polar field strength at the top of the crust, ${C} = 4\pi n_0 e R^2/c$, $R \sim 10$ km  and the electron density is taken to be $n_0 \sim 2.5 \times 10^{34}$ cm$^{-3}$ \citep[eq. 6]{WoodHoll}.\\

For $B_0 = 4 \times 10^{14}$ G, on substitution, we find $t_H = 0.4$ Myr.\\

The above estimate follows from the field evolution study of \citep{WoodHoll,WoodHollPC}, wherein the initial field, $B_0  \sim  4 \times 10^{14}$ G, decays to the final field, $B_{t_H} \sim  0.5 \times B_0$ G, in the Hall timescale $\sim 0.4$ Myr.\\

We can make a simple parameterization in which the magnetic field decays exponentially with time

\begin{equation}
\label{par_B}
B(t) = B_0 e^{-\xi t},
\end{equation}

where the data implies  $\xi = \frac{-\ln(B_{t_H}/B_0)}{C/{B_0}}  =  1.733$ Myr$^{-1}$.\\

The amount of magnetic energy stored in the crust is given roughly by the expression,

\begin{equation}
E_{mag} \sim \frac{B^2}{8\pi}\cdot 4\pi R^2\cdot \Delta R_{cr}~,
\end{equation}

where $B$ is the observed surface polar field,  $R \sim 10^6$ cm is the radius of the star, and $\Delta R_{cr}$ is the thickness
of the crust, which we take to be $\sim 1$ km. The volume of the crust is then, $V =  4\pi R^2\cdot \Delta R_{cr} \sim 4\pi \times 10^{17}$~cm$^{-3}$.\\

In the model, we consider we assume that the magnetic field is decaying with time which implies that the magnetic energy is being released in the process. We also assume that all the loss in magnetic energy is converted into quasi steady thermal emission $L_X$, associated with magnetars.\\

Next, we calculate the X-ray luminosity $L_X $ from the rate of loss of magnetic energy,

\begin{equation}
L_X = \frac{dE_{mag}}{dt} = -B\cdot\frac{dB(t)}{dt}\cdot\frac{V}{4 \pi}  = \xi B_0^2 e^{-2\xi t} \times {10^{17}}~\text{erg s$^{-1}$}
\end{equation}

The X-ray luminosity at the Hall time $t_H $, works out to be approximately

\begin{equation}
L_X \sim 2.2 \times 10^{32}~\text{erg s$^{-1}$}
\end{equation}

Using the simple exponential form of the time dependent magnetic field $B(t)$ above, we can integrate the dipolar spin down equation \ref{dipole} to get $P(t)$ and $\dot P(t)$, whereby

\begin{equation}
P(t) = \sqrt {\frac{10^{-39} B_0^2 \times (1 - e^{-2x}) \times 3.15 \times 10^{13}}{\xi}}~\text{s}
\end{equation}

We find that $P_{t_H} \sim 47$ s when t = $t_H$ and saturates to $P_{t\rightarrow \infty} \sim 54$ s. We note that this dissipation model also shows that star periods saturate, as in the resistive layer model \citep{Rea1}, that will be taken up in the next section. The period saturates at a much higher value compared to that observed for magnetars.\\

We find the instantaneous dipolar spin down radiation energy loss rate for this model, $\dot E $, at the Hall timescale $ t_H $, can be obtained from the period and the magnetic field at this timescale as

\begin{equation}
\dot E \sim  I\omega \cdot\dot\omega\sim\frac{ 4\pi^2 I\times\ 10^{-39}\cdot {B(t_H)}^2}{P^4} \sim 0.33 \times 10^{30}~\text{erg s$^{-1}$}
\end{equation}

Note that at the Hall time scale, $\dot E \ll L_X $. Like $ L_X $, this depends on  $e^{-2\xi t_H}$. For later comparison, we look at the data for this model over a range of spin down ages  We find that

\begin{itemize}
	
	\item For $\tau_{SD} \sim 10$ kyr, we have
	
	\begin{arrowlist}
		\item $P\sim 9.8$ s,
		\item $L_X \sim 8.5 \times 10^{32}$ erg s$^{-1}$, and
		\item $\dot E \sim 7 \times 10^{32} $ erg s$^{-1}$.
	\end{arrowlist}
	
	\item Similarly, for $\tau_{SD} \sim 50$ kyr, we have
	
	\begin{arrowlist}
		\item $P\sim 16.6$ s,
		\item $L_X \sim 7.5 \times 10^{32}$ erg s$^{-1}$, and
		\item $\dot E \sim 7 \times 10^{31} $ erg s$^{-1}$.
	\end{arrowlist}
	
\end{itemize}

Note that at timescales $< 10$ kyr, $\dot E \geq L_X$ erg s$^{-1}$. However after $10$ kyr, $\dot E < L_X$ erg s$^{-1}$.\\

In this model, we get a $L_X$ that is too small in comparison with that observed for magnetars. Also, whereas observations indicate that $L_X $ dies out fast for magnetars after periods $\sim 12$ s, that does not seem to the case here. Actually, $L_X > \dot E$, for large spin down age which is not seen for magnetars or pulsars. We do get period saturation, but at very large periods. Also, from the equation for the period above, for larger fields, $ B_0 > 10^{15} $ G, the period is even larger.

\subsection{Estimation of Magnetic Field Energy Decay Rate Due to Period Saturating Resistive Layer \citep{Rea1}}

One of the puzzles of magnetars is that though binary (accreting) X-ray pulsars, which have smaller magnetic fields are observed at periods of over 30 s, magnetars (till now) have a period that cuts off at around 12 s. It is the copious X ray emission ($> 10^{34} $ erg s$^{-1}$) which is the likely reason for this. \cite{Rea1} have considered a resistive layer in the crust as a way to saturate periods at around those seen in magnetars, $10-20$ s. We saw that the dissipation model in section \ref{sec2.2} also had the feature of period saturation, albeit at a rather large period.\\

In their work, the resistive layer is modelled by an impurity parameter and  starting with an initial magnetic field of median magnetar strength ($B_0 = 3 \times 10^{14 }$ G). This model can achieve a drop in $B$ ($= 2 \times 10^{14}$ G) in $ t \sim 100$ kyr \citep[fig. 3]{Rea1}. At later times there is a simultaneous steep drop in both the magnetic field and $\dot P$.\\

The X-ray luminosity $L_X$, that follows from the rate of loss of magnetic energy  can be estimated from \citep[figs. 3, 4]{Rea1} as

\begin{equation}
L_X = \frac{dE_{mag}}{dt} = \frac{dB(t)^2}{dt} \cdot \frac{V}{8 \pi}
\end{equation}

When the star evolves to a period of $P\sim 10$ s and a spin down age of $\tau_{SD} \sim 80$ kyr, $L_X$ for model A (black line) is then given by $L_X \sim 6 \times 10^{32}$ erg s$^{-1}$.\\

We note that the dipolar spin down radiation energy loss rate for this model $ \dot E $ can also be estimated from \cite[figs. 3, 4]{Rea1} as $\dot E \sim  6 \times 10^{31}$ erg s$^{-1}$.\\

Their results record a steep drop in $B(t)$ and  $\dot P$, after $\tau_{SD} \sim 100$ kyr, so we expect a consequent drop in $L_X$ after that.\\

For later comparison, we can also estimate $L_X$ and $\dot E$ from the figures when the star has evolved to a period of $P \sim 5$ s, where the spin down age $\tau_{SD} \sim  8 $  kyr. At this point in the stars evolution, we find for their model A, $L_X \sim 1.3 \times 10^{34}$ erg s$^{-1}$ and the dipolar spin down radiation energy loss rate is $\dot E \sim 3.2 \times 10^{33}$ erg s$^{-1}$.\\

We thus find that in this model that for  a period, $P > 5$ s and the spin down age, $\tau_{SD} >  8$ kyr,  $\dot E < L_X $.\\

For the mainline magnetars, $L_X $ is generally larger than $\dot E $.
In this model, we do get $L_X $ that is comparable with that observed for magnetars, and we also get period saturation. However, as found before in section \ref{sec2.2}, for large fields of $B_0 > 10^{15}$ G, the period is likely to saturate at values much larger than those observed for magnetars. Also, $L_X$ continues to be greater than $\dot E$ for large spin down age which is not seen for magnetars or high field pulsars. In the following section, we shall consider ohmic dissipation in the presence of large magnetic fields.

\section{The Ohmic Energy Dissipation Time Scale in the Presence of Large Magnetic Fields}
\label{sec3}

The main charge transport in the crust is due to electron currents. In what follows, we write the equations governing electron motion in a medium with electric and magnetic fields. These results are for the single particle equations where the direct  effects of the Fermi sea have not been factored in, but are replaced by an average equilibrium velocity called the drift velocity that  describes the system. Such a situation is often likened to the Drude conductivity. For high polar magnetic fields in the magnetar ball park, $B_p > 10^{14}$ G, the isotropic Fermi sea gets deformed into Landau levels in the direction transverse to the magnetic fields but continues as a one dimensional Fermi sea  along the direction of the magnetic field \citep{ourpreprint}. For example, for large fields, the lowest Landau level has a large degeneracy. Since all the electrons in this level are the same  state, a single particle model for electrons in a magnetic and electric field can be used - the Drude model - to describe the dynamics:

\begin{equation}
m^* \Big(\frac{d\vec v}{dt} +\frac{ \vec v}{\tau}\Big)  =  -e\vec E + \frac{\vec v}{c} x \vec B
\end{equation}

where $\vec v$ is the drift velocity, $ \tau$ is the relaxation time (or collision time) and $m^*$ is the effective mass.\\

In the steady state, $\frac{d\vec v}{dt} = 0 $.\\

The magnetic fields in the crust are determined by currents that run transverse to the magnetic field direction.\\

Let us consider the $ \vec B $ field in the $\vec z$ direction. We can then write the equations in component form,

\begin{align}
v_x & =  -\lambda \Big(E_x +  v_y \frac{B_z}{c}\Big)\\
v_y & =  -\lambda \Big(E_y -  v_x \frac{B_z}{c}\Big)
\end{align}

where $\lambda = e \tau/m^*$, which gives


It is the conductivities in the transverse directions transverse to the magnetic fields that determine the decay of the magnetic fields. It is simple to write these down as

\begin{align}
J_x  &= - n  ev_x  =  n e \frac{\lambda}{\alpha} \Big(E_x - \frac{ \lambda}{c} B_zE_y\Big) \\
J_x &= \sigma_{xx} E_x  + \sigma_{xy} E_y \\
J_y &=  - n e v_y  =  n e \frac{\lambda}{\alpha} \Big(E_y + \frac{ \lambda}{c} B_zE_x\Big)\\
\end{align}

where, $\alpha = (1 + \beta^2 )$ where  $ \beta = \lambda B_z/c$.

We thus find the transverse conductivities as

\begin{align}
\sigma_{xx} & =\sigma_{yy} = \frac{\sigma_0}{\alpha}\\
\sigma_{xy} & = -\sigma_{yx} = -\frac{\sigma_0 \lambda B_z}{c \alpha}
\end{align}
where $\sigma_{xx}$ is the diagonal conductivity and $\sigma_{xy}$ is the non diagonal (Hall) conductivity

and where $\sigma_0 = n e \lambda$ is the isotropic conductivity in the absence of the magnetic field and $\alpha = (1+\beta^2)$, where $ \beta = \lambda B_z/c$. It is to be noted that the transverse conductivities depend on the magnetic field $B_z$, whereas the isotropic conductivity does not.\\

We can now write down the Ohmic magnetic field decay time for the star as

\begin{equation}
\tau^{D}_{ohm} = \frac{ 4\pi\sigma_0L^2}{c^2 \alpha},
\end{equation}

where $\tau^{D}_{ohm} $ is the diagonal transverse conductivity decay time.\\

From the above, we find that the conductivity tensor is highly anisotropic. The parameter that controls the conductivities is $ \beta = \lambda B_z/c$. If this parameter is less than 1, then the conductivities are essentially the isotropic one in the absence of magnetic field, $\sigma_0$. If however, this parameter is greater than 1, then, we can neglect the factor of unity in $\alpha = (1 + \beta^2 )$ and the conductivity becomes magnetic field dependent. For large fields it is the second term in $\alpha = (1 + \beta^2)$, that dominates, eg., $\sigma_{xx} = \sigma_{yy} = \sigma_0/\alpha$ go inversely as the square of the magnetic field.\\

Let us see what is the value of this parameter for some typical values in the crust.\\

We can express $\lambda = \sigma_0/(ne)$, where, $n$ is the electron density and $e$ is the electric charge. For a typical electron density of $ n = 2.5 \times 10^{34}$ (in cgs units) and $\sigma_0 \sim 10^{23}$/s \citep{ChamHaen}, $\alpha \sim\beta^2 \sim 10^{-25} B_{z}^{0}$. It is then evident that for magnetic fields in excess of $ 10^{13}$ G that $\beta > 1$. This is the case for magnetars. We thus have

\begin{equation}
\tau^{D}_{ohm} = \frac{ 4\pi\sigma_0\cdot {L^2}}{ c^2  \beta^2 } = \frac{ 4\pi(n^2 e^2)\cdot {L^2}}{ \sigma_0 B^2},
\end{equation}

where $\tau^{D}_{ohm}$ is the diagonal transverse conductivity decay time.\\

Let us look at ohmic time scale for the polar magnetic field decay in a magnetar which has the canonical value $B = 4 \times 10^{14}$ G.\\

Using the usual inputs for $n = 2.5 \times 10^{34}$ (cgs) \citep{ WoodHoll} above and the length scale transverse to the magnetic field, $L = 10$ km, and $\sigma_0 \sim 10^{23}$ (cgs), we find

\begin{equation}
\tau^{D}_{ohm} \sim 3.3~\text{kyr}
\end{equation}

This translates to $\xi \sim 1/\tau^{D}_{ohm} \sim 300$ Myr$^{-1}$. Note this is much larger than the value for the Hall dissipation, $\xi \sim 1.733$ Myr$^{-1}$. However, as can be seen from the above expression, its actual value will depend on the specific electron density and the conductivity. The above values give a typical range.\\

It is interesting to compare the ohmic decay time scale $\tau^{D}_{ohm}$ with the spin down age as they both depend inversely on ${B}^2$. We have

\begin{equation}
\tau_{SD} = \frac{P}{2\dot P} = 10^{39}\ \frac{P^2}{2 B^2}
\end{equation}

Taking $P \sim 8$ s (from magnetar data) and $B  = 4 \times 10^{14}$ G, it indicates that the spin down age and the Ohmic time scale above are similar!

\section{Dynamo Model ($B_{polar}, L_X$)}
\label{sec4}

In the dynamo model, all fields are created at birth. They undergo evolution by ambipolar diffusion in the core interior which is sustained by $n, p, e, \beta$ equilibrium. In the crust, there are no free protons and neutrons (no $\beta$ equilibrium) and dissipation occurs via Hall effect field line reconnection and Ohmic dissipation. The polar field $B(t)$ can only go down with time as magnetic energy is dissipated. Thus, we expect a monotonic decrease of the magnetic field throughout the star. The ambipolar dissipation in the core and the ohmic dissipation in the crust both gives rise to X-ray emission. For the dynamo model, the polar surface magnetic field and X-ray luminosity are largest at birth and then start falling as the magnetic energy gets dissipated.\\

If the crustal fields are high enough, then the magnetic field dependent transverse conductivities come into play \citep{Israeli} and  the `crystal' is quasi filamentary at birth -- screening length less than the inter atomic distance \citep{ShaRed,Bed,ourpreprint}. We have found \citep{ourpreprint} that if the matter density is below a certain threshold in the outer crust ($\sim 10^{7-8}$ g cm$^{-3}$), the Landau radius is very much smaller than half the inter-atomic distance, confining the electrons in the lowest Landau level. Furthermore, the Coulomb interaction between the ions and the electrons pulls all the electrons into neutral cylindrical atoms. This implies that whereas the whole crust is a filamentary `crystal', the outer crust is additionally an insulator for magnetar strength fields ($B \sim 10^{15 - 14}$ G). Thus, we are not likely to have magnetic strains that produce conventional crystal cleaving in this region of the crust (this suppresses glitches/flares) at the beginning when the fields are the largest.\\

We have also found that the decay of the magnetic field can be different for different models, depending on the resistivity, which in turn can depend on the magnetic field. Given the resistivity for typical initial fields considered in the cases above ($B_0 \sim 3-4 \times 10 ^{14}$ G), it is instructive to make a simple parameterization in which the instantaneous polar (crustal) magnetic field, $B(t)$, decays exponentially with a constant decay constant, $\xi \sim 1/\tau^D_{ohm}$.\\

As a baseline, we shall work with a typical timescale for magnetar of fields, $B_0 = 4 \times 10 ^{14}$ G set by the considerations in section \ref{sec3}. Hence we have,

\begin{equation}
\xi \sim \frac{1}{\tau^{D}_{ohm}} \sim 300~\text{Myr$^{-1}$}
\end{equation}

Furthermore, we use the parameterization that was employed in section \ref{sec2.2}, ie.,

\begin{equation*}
B(t) = B_0 e^{-\xi t}
\end{equation*}

Using the simple exponential form of the time dependent magnetic field $B(t)$ above, we can integrate the dipolar spin down equation to get $ P(t)$ and  $\dot P(t)$ as

\begin{equation}
\label{period}
P(t) \times \dot P(t)  = 10^{-39} B(t)^2
\end{equation}

which yields

\begin{equation}
P(t)^2 - P(0)^2 = 10^{-39} B(t)^2 \times \frac{e^{2x} - 1}{\xi} = 10^{-39} B_0^2 \times \frac{1 - e^{-2x}}{\xi},
\end{equation}

where $x = \xi t $ and $\xi$ is be expressed in units of Myr$^{-1}$ from the expressions as in section \ref{sec3}.\\

Here, $P(0)$ is the initial period at birth, and can be neglected. The initial period for magnetars is expected to be a few milliseconds, so we may start our evolution at $P \sim 100$ ms.\\

Thus,

\begin{align}
P(t) &= \sqrt{10^{-39} B(t)^2 \times \frac{e^{2x} - 1}{\xi} \times 3.15 \times 10^{13}}\\
\implies P(t) &= \sqrt {10^{-39} B_0^2 \times \frac{1 - e^{-2x}}{\xi} \times 3.15 \times 10^{13}}
\end{align}

Once we have this simple expression for $P(t)$, we can get $\dot P(t)$ from eq. \ref{period}. So,

\begin{equation}
\dot P(t) = 10^{-39} \times \frac{B(t)^2}{P(t)} = 10^{-39} B_0^2 \times \frac{e^{-2x}}{P(t)}
\end{equation}

We have deliberately written the period in terms of both $B(t)$ and the initial field $B_0$. The former is to make contact with observations and the latter form shows that for any initial or starting field, there is a maximum period in this simplified parameterization as observed in \cite{Rea1} -- a matter of interest as there is a cap in the observation of magnetar periods. Given an initial field $ B_0$, we can find the maximum period in this model by taking the limit ${x\rightarrow\infty}$, which is given by

\begin{equation}
P^{max} ({t\rightarrow\infty}) = \sqrt {10^{-39}\times \frac{B_0^2}{\xi}}~\text{s}
\end{equation}

For illustration, for the case of $B_0 = 4 \times 10^{14}$ G and a typical $\xi = 300$ Myr$^{-1}$, $P^{max} \sim 4.1$ s.\\

The magnetar catalogue \citep{catalog} suggests that this is too low. However, as is evident from above, $P^{max}$ is proportional not only to $B_0$ but goes inversely as $\xi$. Though we have taken $\xi$ to be constant, we know from  section \ref{sec3} that it depends on $B(t)$, and so goes down with time in this case. This would no longer leave the period constant -- $P$ would keep increasing. Moreover, for the high field magnetar SGR 1806-20, which has $B_0$ greater than the observed $B(t) \sim 2 \times 10^{15}$ G, $P$ would be over five times larger. This is somewhat higher than the observed $ P^{max}$ for magnetars.\\


We have assumed a dissipative mechanism where the polar magnetic field falls  as $B(t) = B_0 e^{-\xi t}$. We find that $\dot P(t)$ falls even more sharply as $e^{-2{\xi t}}$, which is consistent with the resistive layer scenario.\\

It is interesting to note that for a given $\xi$, the spin down (characteristic) age,

\begin{equation}
\tau_{SD}^{Myr} = \frac{P(t)}{2\dot P(t)} = \frac{(e^{2x} - 1)}{2~\xi},
\end{equation}
 
is independent of the magnetic field in this model. Therefore, given a $ \tau_{SD} $, we can fix the value of $x$ in this model and then determine the period and other parameters.\\



Note that in this model, the period can be written as

\begin{equation}
P(t) = \sqrt {10^{-39} B(t)^2 \times 2~\tau_{SD}}
\end{equation}

We can calculate $B_0$, the final field at the surface for each magnetar from the assumed dependence of the surface B(t), ie., $B(t) = B_0~e^{-x} $.

\subsection{X-Ray Luminosity ($L_X$)}
\label{sec4.1}

The X-ray luminosity $ L_X $  is made up of the sum of the X-ray luminosity from the core, $L_{A,X}$ which is transported out by ambipolar diffusion and the crustal luminosity and $L_{ohm}$, which arises from the dissipation of the currents in the crust. This is because there is hardly any resistive dissipation in the core as the conductivity is very high. Since we do not have a microscopic or even a macroscopic description of the magnetohydrodynamics of the star interior, we shall make some heuristic estimates of a rather complex system. This is with a view to draw out some features of the model.

\subsection{Crustal Ohmic Dissipation}
\label{sec4.2}

In the interest of simplicity, we shall assume that that $L_X$ arises solely from Ohmic dissipation in the crust, as in the models in section \ref{sec1}, and that the surface polar field is due to crustal currents. We further take $\xi \sim 1/\tau^{D}_{ohm}$ to be constant. We will comment later that neither of these simplifications disturb the features we uncover. Given these expressions for the instantaneous period, its derivative and the magnetic field, we can write down the expression for $L_{ohm}$ as

\begin{equation}
L_{ohm} = 2B(t)\frac{dB(t)}{dt} \cdot \frac{\text{Vol}}{8 \pi}
\end{equation}

\begin{equation}
\implies L_{ohm} = \Big(\frac{\xi}{3.15 \times 10^{13}}\Big) B(t)^2 \times 10 ^{17}  =  \Big(\frac{\xi}{3.15 \times 10^{13}}\Big) B_0^2~e^{-2x} \times {10^{17}}~\text{erg s$^{-1}$,}
\end{equation}

where Vol $= 4\pi R^2\cdot \Delta R_{cr}$
is the volume of the crust, which we take to be of radius, $R\sim 10$ km and a thickness, $\Delta R_{cr} \sim 1$ km. Since we know $L_{ohm}$ and $B(t)$, we can determine $\xi$.\\

We find that the values of $\xi$ found from the above consideration (see table \ref{dyn-table}) do not match with those estimated in section \ref{sec3}. Recall that $\xi = 1/\tau^{D}_{ohm}$ for $B \sim 4 \times 10^{14}$ G and $\sigma_0 \sim 10^{23}$ (CGS), $\xi \sim 300$ Myr$^{-1}$ and $\xi$ is proportional to $B(t)^2$.

\subsubsection*{Examples}

We shall present estimates for three different magnetars.

\begin{enumerate}
	\item 4U 0142 + 61, whose spin down age is large $\sim 0.068$ Myr, small surface polar field $B(t) \sim 1.34 \times 10^{14}$ G but large $L_{X}^{thermal} \sim 0.25 \times 10^{35}$ erg s$^{-1}$.
	
	\item CXOU J010043, a typical magnetar in the intermediate range, whose spin down age is $\sim 0.0068$ Myr, surface polar field $B(t) \sim 3.9 \times 10^{14}$ G, and  $L_{X}^{thermal} \sim 0.65 \times 10^{35}$ erg s$^{-1}$.
	
	\item SGR 1806, whose spin down age is small $\sim 0.00025$ My, large surface polar field, $B(t) \sim 20 \times 10^{14} $G, and $L_{X}^{thermal} \sim 1.63 \times 10^{35}$ erg s$^{-1}$.
	
\end{enumerate}

The dipolar spin down radiation energy loss rate for this model $\dot E$ can be determined from the period and the instantaneous magnetic field,

\begin{equation}
\dot E = \frac{4 \pi^2 I\dot P}{P^3}~\text{erg s$^{-1}$}
\end{equation}
\\
and thus we can also determine their ratio which is noteworthy for magnetars. The figures indicate that, $\dot E/L_X $ remains small all the way to $\tau_{SD} \sim 1$ Myr.\\

Note that both $L_{ohm}$ and $\dot E$ go down exponentially as $e^{-2x}$ with time. The table and figures in the next section are a summary of our results\\

\subsection*{Remarks \& Plots}

A puzzling feature of the  magnetar SGR 1806, which has a large surface polar field $ B (t) \sim 20 \times 10^{14} $ G is the small values of $\xi \sim 13$ Myr$^{-1}$ and $x$. If $\xi \sim B_{polar}^2 $ as found in section \ref{sec3}, it would yield $\xi \sim 7500 $ Myr$^{-1}$ at an electron density $n_e \sim 2.5 \times 10^{34}$ g cm$^{-3}$. This is over $500$ times larger than the one required to fit the data. The data seems to imply that Ohmic decay is much slower than expected. A physical reason can be that the Ohmic decay in this model happens at higher density in the inner crust -- recall that $1/\xi = \tau^{D}_{ohm} = \frac{4\pi({n_e}^2 e^2){L^2}}{\sigma_0 B^2 }$.\\

Another significant point is that in this work, we have not included the dissipation from ambipolar diffusion. This shall be included in a coming work. Our preliminary finding indicates that the late time behaviour is similar in both cases, that is, $\dot E/L_X$ continues to be $<<1$ over ages that go to a million years. 


\begin{center}
\begin{table}[ht]
	\centering
	\caption{Dynamo Model Values}
	\label{dyn-table}
	\begin{tabular}{lllll}
		\hline
		Magnetar & $x$ & $\xi$ (Myr$^{-1}$) & $t/\tau$ & $B_0$ (G)\\ \hline
		4U0142+61 & 2.052 & 438.572 & 0.068 & $1.04 \times 10^{15}$\\
		CXOU J010043.1-721134 & 0.513 & 132.568 & 0.572 & $6.56 \times 10^{14}$\\
		SGR 1806-20 & 0.003 & 13.1195 & 0.945 & $1.96 \times 10^{15}$\\ \hline
	\end{tabular}
\end{table}
\end{center}

\begin{figure}[H]
	\centering
	\begin{minipage}[b]{0.4\textwidth}
		\includegraphics[scale=0.45]{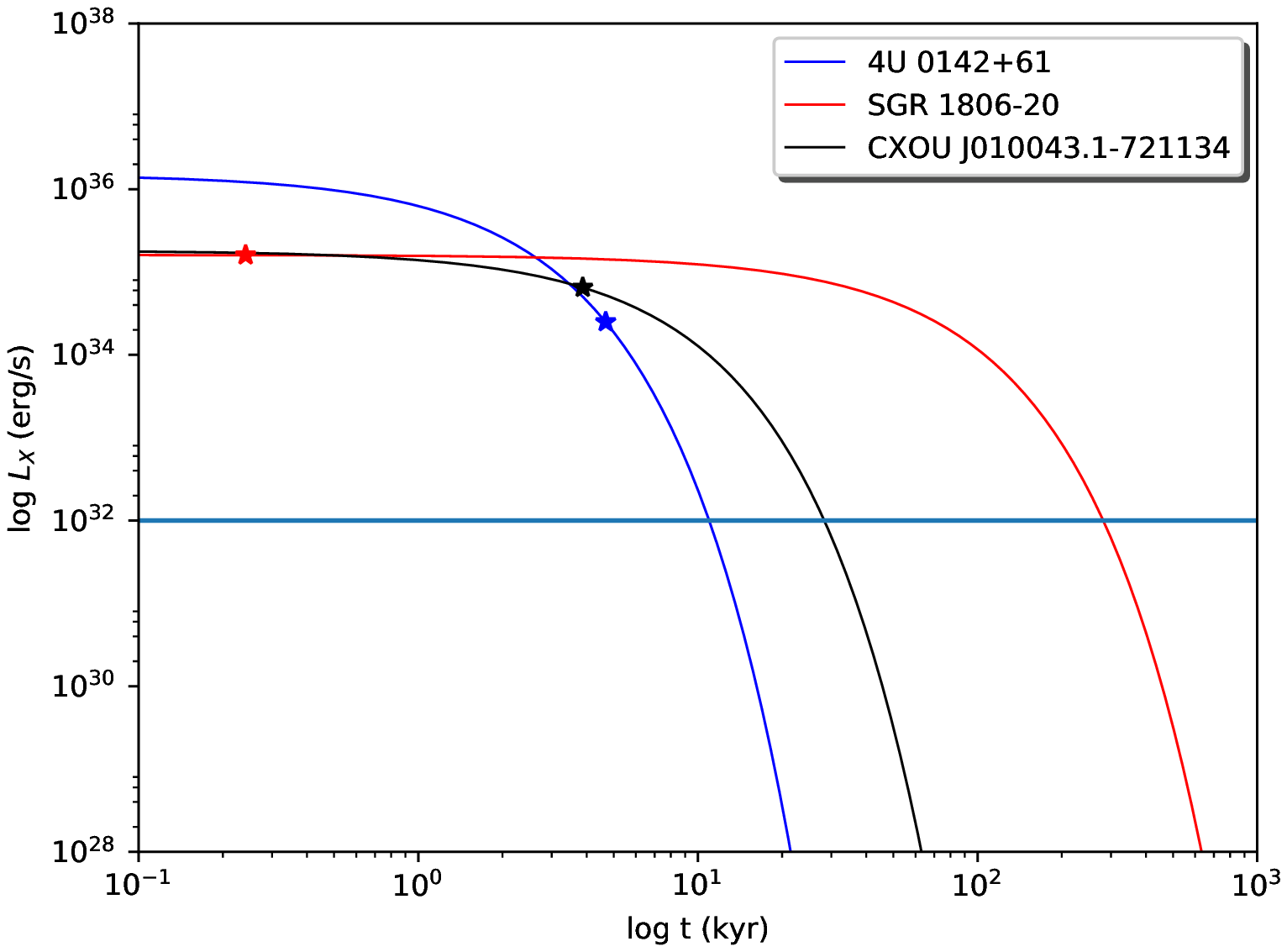}
		\caption{\footnotesize X-ray Luminosity Vs. Time}
	\end{minipage}
	\hfill
		\begin{minipage}[b]{0.4\textwidth}
		\includegraphics[scale=0.45]{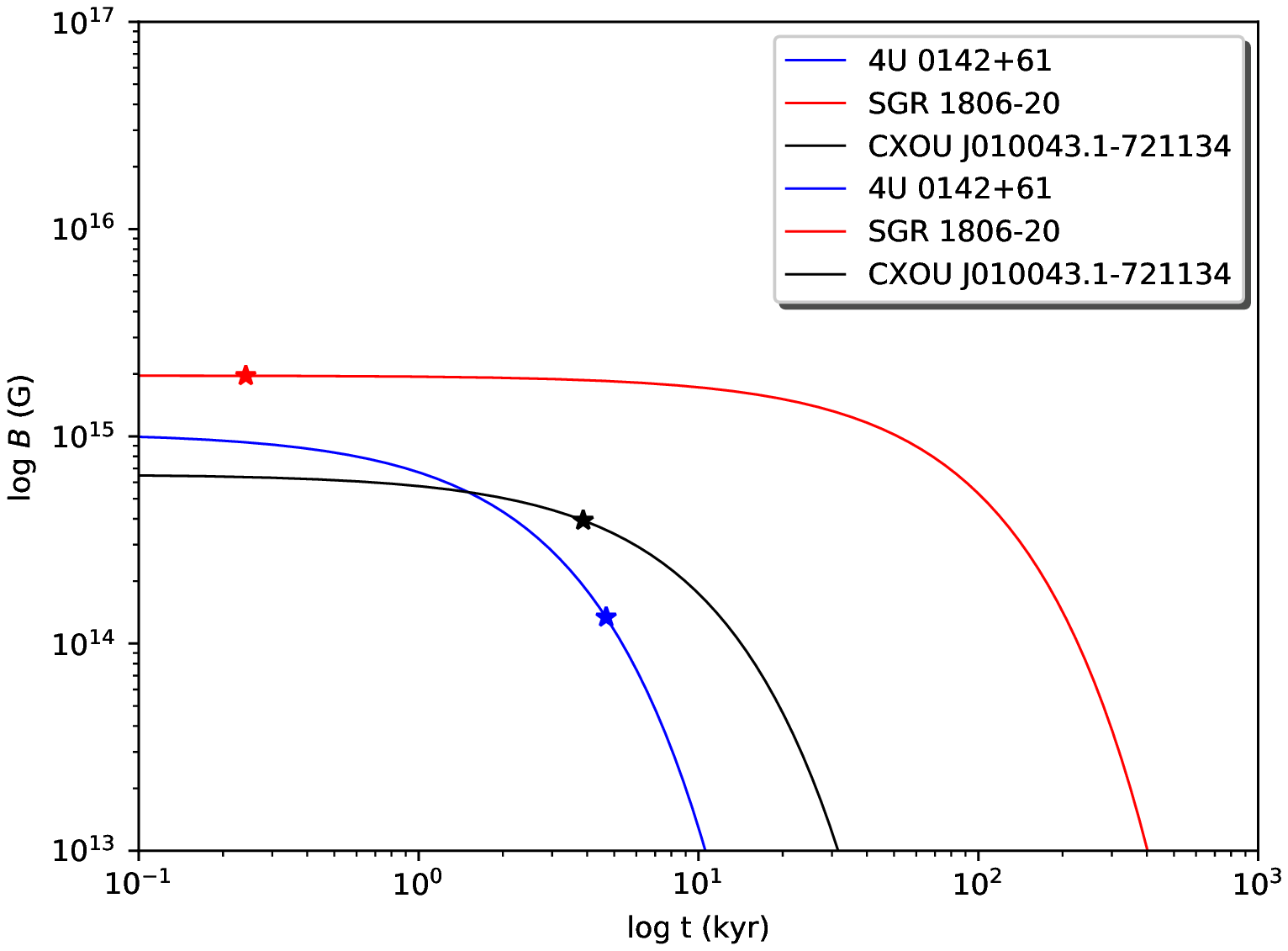}
		\caption{\footnotesize $B(t)$ Vs. Time}
	\end{minipage}

\end{figure}

\begin{figure}[H]
	\centering
		\begin{minipage}[b]{0.4\textwidth}
		\includegraphics[scale=0.45]{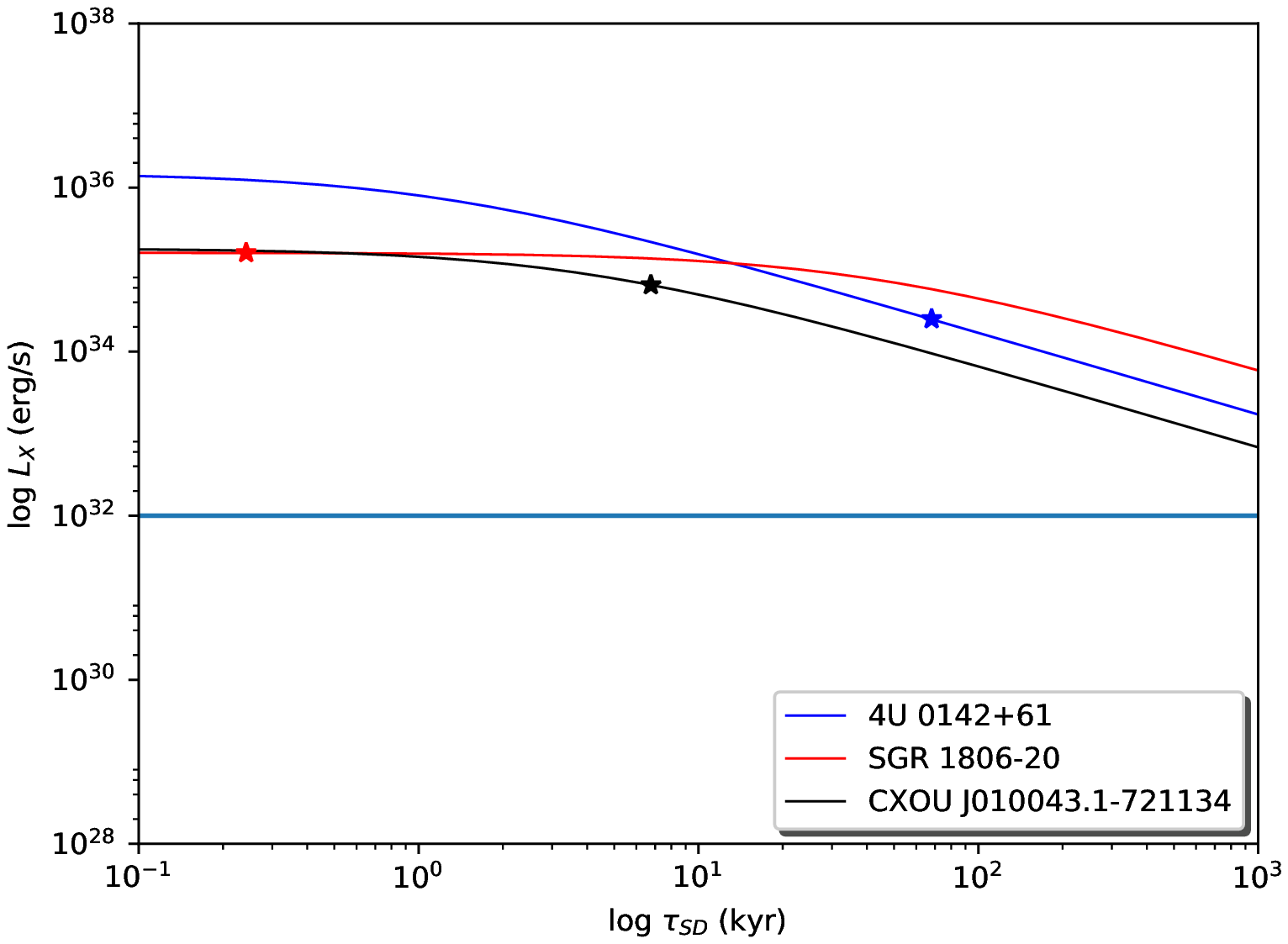}
		\caption{\footnotesize X-ray Luminosity Vs. Spin Down Age}
	\end{minipage}
	\hfill
	\begin{minipage}[b]{0.4\textwidth}
		\includegraphics[scale=0.45]{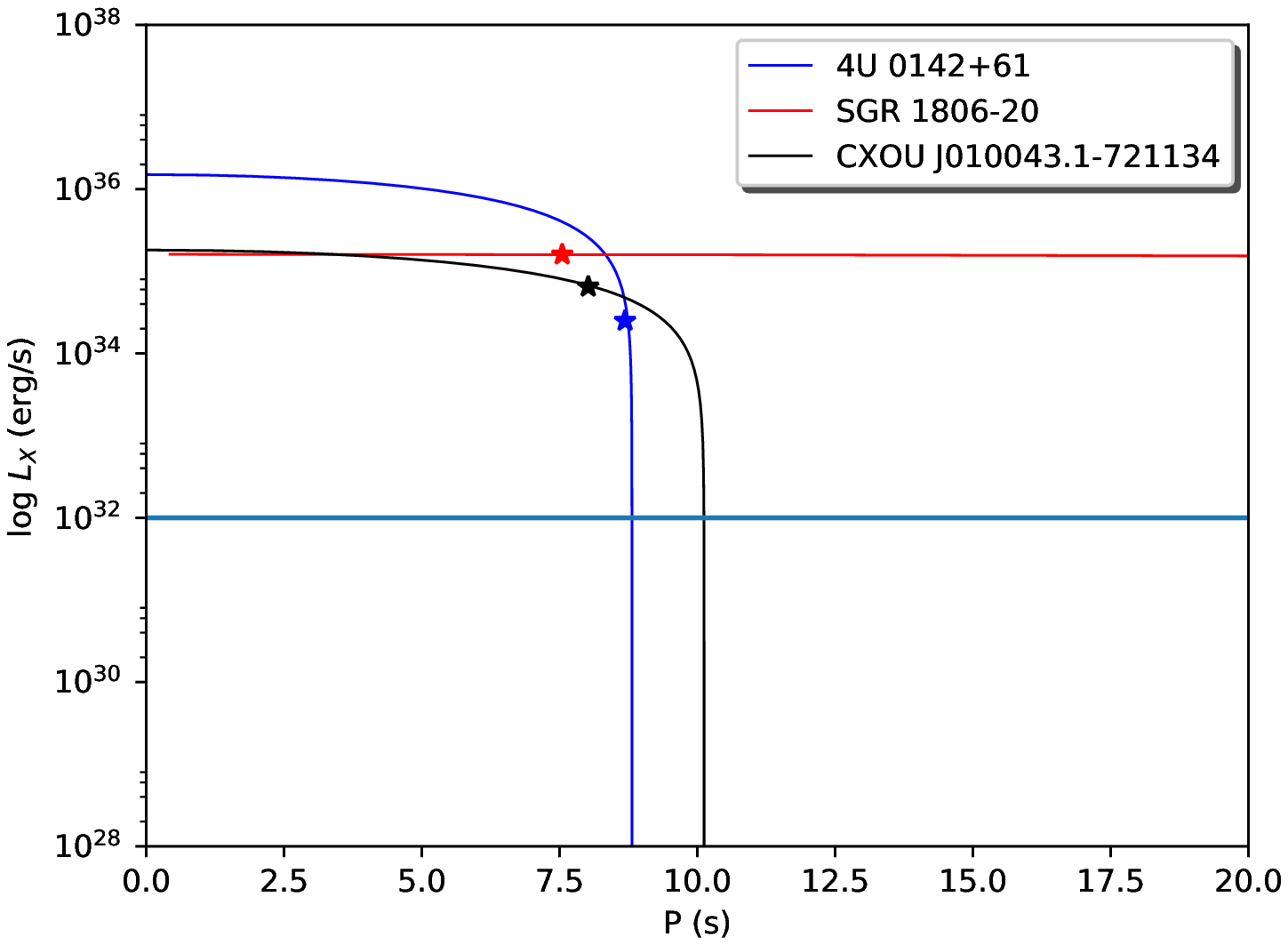}
		\caption{\footnotesize X-ray Luminosity Vs. Period}
	\end{minipage}

\end{figure}

\begin{figure}[H]
	\centering
	\begin{minipage}[b]{0.4\textwidth}
		\includegraphics[scale=0.45]{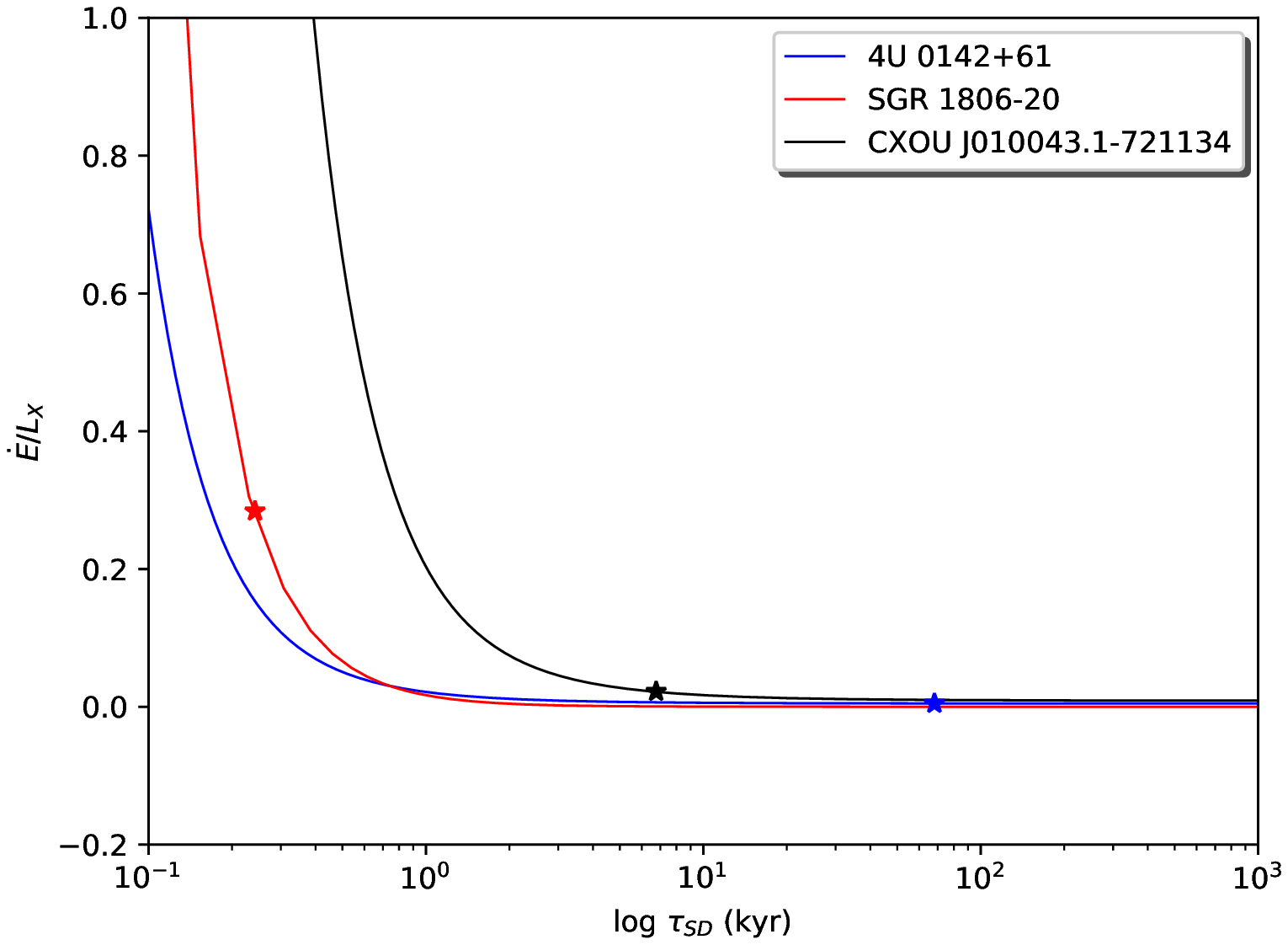}
		\caption{\footnotesize $\dot E/L_X$ Vs. Spin Down Age}
\end{minipage}
	\hfill
	\begin{minipage}[b]{0.4\textwidth}
		\includegraphics[scale=0.45]{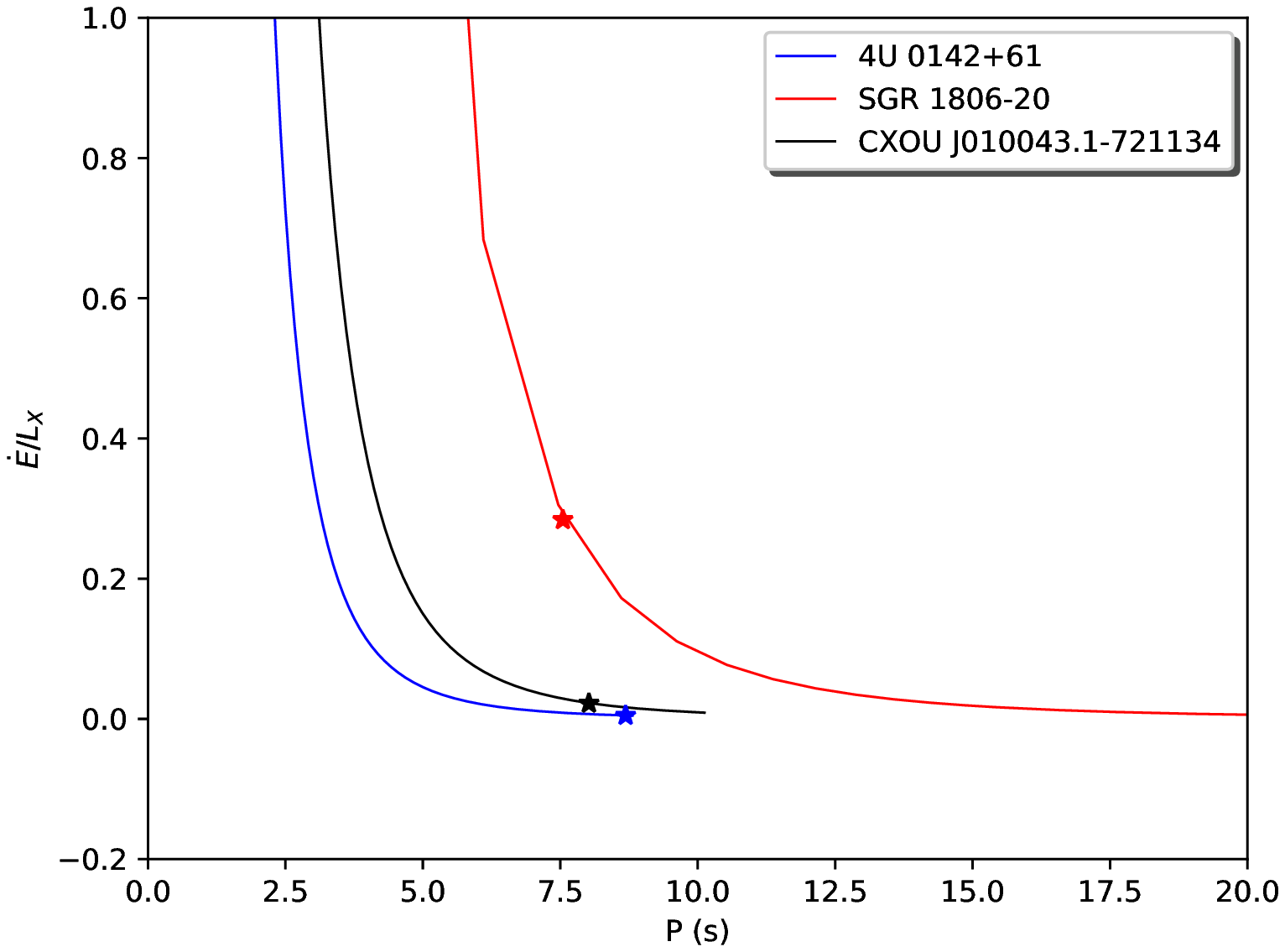}
		\caption{\footnotesize $\dot E/L_X$ Vs. Period}
	\end{minipage}
\end{figure}

The horizontal line in the plots with X-ray luminosity corresponds to a value of $L_X = 10^{32}$ erg s$^{-1}$. This bound is marked due to magnetars becoming potentially difficult to be observed below this value.

\section{The Screened Core Model $B_{polar}, L_x$}

In the regular dynamo model, it is the currents created at birth that produce the magnetic field over the entire star. It is the dissipation of these currents that gives rise to field decay and X-ray radiation. On the other hand, in the screened core model, the core field is dynamically created at birth by the alignment of neutron spins by the strong interaction ground state -- it does not arise from currents. However, the plasma around the flux created by this dynamical field generation give rise to screening currents that shield the core field \citep{DipSoni,mag} and stop it from spreading out. The screening currents have to not only neutralise the core field at the outer core and the crust but also turn back and  compress the field lines into the volume inside. These currents thus carry an energy that is much higher than a regular dipolar field. We shall comment on this below. Thus in this model, it is the screening currents that dissipate allowing the core field out in its relaxed configuration.\\

For the dynamo model crustal, currents dissipation decreases the polar and core magnetic fields, whereas in this model, with the dissipation of screening currents, the field goes up. Thus, the final configuration here is rather different. Also, the energy loss (which depends on the ambient magnetic field) will not be the same as in the dynamo model, though the dissipation mechanisms continue to be ambipolar diffusion and Ohmic dissipation.\\

In the screened core model, the screening currents that shield the dynamical core field are first dissipated by ambipolar diffusion in the outer core and transported out to the crust. Ohmic dissipation and Hall conductance then kick in and the core field rises to its unscreened value at the surface and then stays constant.\\

When $B(t)$ in the crust exceeds $ 10^{13-14}$ G, Landau Radius $ < $ Bohr radius. Furthermore, the crustal crystal cleaves when the  magnetic field difference between the inner and outer crust exceeds $B(t) > 10^{13-14}$ G with accompanying glitches or flares. As screening currents dissipate and $B(t)$ goes further up, the transverse (to the polar magnetic field) conductivities become field dependent and diminish, resulting in a quasi-filamentary crystal in the direction of $B(t)$. Finally, as the inter ionic distance becomes very much larger than the Landau radius, the ions can attract all the electrons to form cylindrically deformed neutral atoms and the outer crust can become an insulator. This sequence of events is just the reverse of the dynamo model.\\

This case is the opposite of the previous dynamo model where the surface polar field decays exponentially. In this case, the surface polar field goes up with time till the screening currents dissipate completely. We choose a particular parameterization where polar magnetic field which increases with time,

\begin{equation}
B(t) = B_0 \sqrt{1-e^{-\xi t}}
\end{equation}

This somewhat mysterious form of the above parameterization for the time dependence of the magnetic field evolution becomes necessary if we want to accommodate the time varying magnetic field dependence of $\xi$. As we had pointed out in section \ref{sec3}, $\xi$ is approximately proportional to $ B(t)^2$ via the conductivity. Since this field is time dependent and depends on $\xi t$ (above), we need to establish the existence of a solution for the parameterization of $B(t)$ for all $t$ above. If we use the form above without the square root, we do not have a solution for the case of a time varying $\xi(t)$ that can satisfy the condition that the solution is well behaved at $t \sim 0$.\\

However, as in section \ref{sec4}, we shall work with a constant $\xi$. Using the above form of the time dependent magnetic field $B(t)$ above, we can integrate the dipolar spin down eq. \ref{dipole} to get $P(t)$ and $\dot P(t)$ as

\begin{center}
$P(t)\cdot \dot P(t) = 10^{-39} B(t)^2$
\end{center}

\begin{gather}
\implies P(t)^2 - P(0)^2 = 10^{-39} B_0^2 \times \frac{2F(x)}{\xi} \times 3.15 \times 10^{13}\\
\text{where}~F({x}) =  (x -1 + e^{-x})
\end{gather}

Here, $x = \xi t$ and $\xi$ is expressed in units of Myr$^{-1}$ as in section \ref{sec4}. Also, as before, $P(0)$ is the initial period at birth and can be neglected.\\

Thus, 

\begin{align}
P(t) &= \sqrt {10^{-39} B_0^2 \times \frac{2 F(x)}{\xi} \times 3.15 \times 10^{13}}\\
\implies P(t) &= \sqrt {10^{-39} B(t)^2 \times \frac{2 F(x)}{\xi (1 - e^{-x})} \times 3.15 \times 10^{13}}
\end{align}

Once we have this simple expression for $P(t)$, we can get $\dot P(t)$ from the first equation,

\begin{equation*}
\dot P(t) =  10^{-39} \frac{B(t)^2}{P(t)} = 10^{-39} B_0^2 \times \frac{1 - e^{-x}}{P(t)}
\end{equation*}

Given an initial field $B_0$, we can find the time period at late times in this model by taking the limit $x \gg 1$, which is given by

\begin{equation*}
P (t\rightarrow\infty) = \sqrt {10^{-39} B_0^2 \frac{2 F(x)}{\xi} \times 3.15 \times 10^{13}}
\end{equation*}

This shows that unlike the previous case, there is no cap on the period which goes as $x^{1/2}$ at late times. In this scenario, the late time evolution of magnetars shows $\dot P(t)$, falls off as $x^{-1/2}$, which is not inconsistent with the observed values. The characteristic age is,


\begin{equation}
\label{eq_tau}
\tau_{SD}^{Myr} = \frac{P(t)}{2\dot P(t)} = \frac{ F(x)}{ \xi (1-e^{-x})} 	
\end{equation}

Since $\tau_{SD}$ is known from the observations \citep{catalog}, if we know the value of $\xi$, we can fix the value of $x$ in this model and then determine the period and other parameters.\\

Notice that in this model, the period can also be written as $P(t) = \sqrt{10^{-39} B(t)^2 \times 2 \tau_{SD}}$


\subsection{Crustal Ohmic Dissipation}

The X-ray luminosity $L_X$ is made up of the sum of the X-ray luminosity from the core $L_{A,X}$, which is transported out by ambipolar diffusion and the crustal luminosity $L_{ohm}$, which arises from the dissipation of the screening currents in the crust.\\

It is the screening field $B_s(t)$ in the crust which dissipates to generate $L_{ohm}$. We make the assumption that the crustal field approximates to the surface field $B(t)$. The screening field is given by

\begin{equation}
B_s = B_0 - B(t) = B_0~(1 - \sqrt{ 1- e^{-x}})
\end{equation}

$B_0$ can be determined from the observed instantaneous value of $B(t)$ provided we know the instantaneous value of $x$. The expression for, $L_{ohm}$ that follows is,

\begin{equation}
L_{ohm}  = \frac{dB_s(t)^2}{dt} \cdot \frac{\text{Vol}}{8 \pi} = \frac{1}{2}~B(t)^2 \Big(\frac{\xi}{3.15 \times 10^{13}}\Big) \frac{e^{-x} [(1-e^{-x})^{1/2}-1]}{ (1-e^{-x})^{3/2}} \cdot {10^{17}}~ \text{erg s$^{-1}$}
\end{equation}

where Vol $ = 4\pi R^2\cdot \Delta R_{cr}$ is the volume of the crust. Like in the dynamo model, we take the radius $R\sim 10$ km and a thickness $\Delta R_{cr} \sim 1$ km.\\

In this model, $L_{ohm}$ is a function not only $\xi$, but also of $x$. We can thus express $\xi $ in terms of $L_X$ and $x$ from this relation. On equating the value of $\xi$ obtained from eq. \ref{eq_tau}, we get

\begin{equation}
\label{L_ohm}
L_{ohm} =  B_{14}(t)^2 \frac {10^{35}}{6.3~\tau_{SD}^{kyr}} \times \frac{ F(x) \cdot e^{-x} [(1-e^{-x})^{1/2}-1]}{ (1-e^{-x})^{5/2}}~\text{erg s$^{-1}$}
\end{equation}

where $\tau_{SD}^{kyr}$ is expressed in kyr and $B_{14}$ is the magnetic field in units of $10^{14}$ G.\\

Unlike for the dynamo model, we find that for magnetars 4U 0142+61 and CXOU J010043.1-721134, the observed $B_{14}(t)^2 $ and $\tau_{SD}^{kyr}$ put an upper bound for $L_{ohm}$ that falls well short of the observed $L_X$. We then have to include the ambipolar diffusion contribution from the outer core to explain the data, which we do in what follows.\\

Furthermore, here we cannot determine $\xi $ from  $L_X$ as for the dynamo model, so we use the results of section \ref{sec3}, where $\xi$ was found to be taken to be proportional to $B(t)^2$. By scaling the baseline value of $\xi $ according to the relevant $B(t)$, we can estimate the value of $\xi$ for different magnetars. We can then use the equation for $\tau_{SD}$ (eq. \ref{eq_tau}) to determine $x$ and $B_0$.\\

On the other hand for the magnetar SGR 1806-20, the bound is much larger than the observed $L_X $ and so can be eminently accommodated by $L_{ohm}$. Thus, SGR 1806-20 is the only magnetar in the screened core model (our model), where we can follow the steps used for the dynamo model and approximate $L_X \sim L_{ohm}$. The results and evolutionary track of this magnetar is given in the table and figures that follow.

\subsection{Ambipolar Diffusion}

To explain the $L_X$ for the magnetars 4U 0142 + 61 and CXOU J010043.1-721134, we need to invoke ambipolar diffusion, which is a dissipative mechanism that operates in the high temperature ($T \sim  10^{8-9} $ K) inner region of the star -- the core and the outer core which has free electrons, neutrons and protons. Ambipolar diffusion transports the the magnetic field outwards to the crust as the screening currents dissipate. This is in contrast to Ohmic dissipation which operates only in the crust where the conductivity is not so high.\\

The screening currents lock in and squeeze the core field in the region of the outer core. The outer core is the region between the magnetized inner core and the crust. We use an averaged flux conservation to approximate the magnetic field configuration in this region. As the screening currents dissipate and move radially out, the magnetic energy of the squeezed configuration will diminish. Since $L_{ohm} $ is very small in this case, we simplify our treatment by approximating the entire contribution to $L_X $ to be from the ambipolar outer core screening currents. In this case, the final unscreened field at the surface must be identified with the free field value of the dipole core field. The core field at the radius $r_c$, which is at the surface of the inner core is assumed to arise from a uniform dipole moment density and maintain a value of $B_c = 10^{16} $ G for all magnetars. This approximation is plausible as \cite{Haensel} finds that the core density varies only  slightly with the stellar mass. Since different magnetars have different surface fields, the free field value of each magnetar at the surface $B_0$ is set by matching the value of $r_c$ for each magnetar, where the subscript $c$ denotes the parameter at the inner core.\\

First, we need to calculate $B_0 $, the final field at the surface for each star. We have

\begin{gather*}
B(t) =  B_0 \sqrt {1 - e^ {-x} }~\text{(from the parameterization), and}\\
\tau_{SD}^{Myr} = \frac{F(x)}{\xi (1 - e^{-x})}~\text{(spin down age)}
\end{gather*}

We assume that the crustal screening field is $B_0$ and we can determine  $\xi (B_0)$ by scaling with $ B_0 $ as

\begin{equation}
\xi (B_0) = \xi_I \times \frac{B_0^2}{B_I^2} = \xi_I \times \frac{B(t)^2}{B_I^2 (1 - e^{-x})} 
\end{equation}

where $\xi_I$ and $B_I^2$  are some standard initial values from section \ref{sec3}. Hence,

\begin{equation}
\tau_{SD} = F(x) \times \frac{B_I^2}{\xi_I B(t)^2}
\end{equation}

Since the instantaneous spin down age and $B(t)$ are known from the data \citep{catalog}, we can determine $x$ and then $B_0$. Now, if we fix $B_{core} = 10^{16}$ G,  then, $B_0 = B_{core} (r_c/R)^3$, where we have taken both the surface and crust radius to be, R. This will determine $r_c$ for each magnetar so it matches with the final $B_0$.\\

Next, we move on to pinning down the magnetic energy locked in the screening currents for which  write the `averaged' flux conservation condition,

\begin{equation}\label{eq1}
(r^2- r_c^2) B_r = r_c^2 B_c
\end{equation}

To avoid the singularity at $r = r_c$ in $L_X$ (ambipolar), we have moved our starting point to $r = 2 r_c$ at $t = 0$. We can now use the results of \cite{Passamonti} for the magnetic field transport,

\begin{equation}
\label{r_time_dep}
r(t) = 2 r_c\ e^{t/\tau_B}
\end{equation}

where $\tau_B$ is the ambipolar timescale and $1/\tau_B = \xi_2$. $R$ can be taken to be $10$ km at the crust.\\

Next, we determine the time $t = t_r$ for $r(t)$ to reach $r > 2 r_c$

\begin{equation}
\label{t_r}
t_r = \frac{1}{\xi_2}~ln \Big(\frac{r}{2r_c}\Big)
\end{equation}

Let us define the unscreened $B$-field at a radius $r$ in the outer core to be $B_{r}^{0}$. Then, 

\begin{equation}
B_{r}^{0} = B_c \Big(\frac{r_c}{r}\Big)^3
\end{equation}

The induced field due to the screening currents, $B_s(r)$ (screening field) that dissipates is,

\begin{equation}
B_s(r) = (B_{r}^{0} - B_r)
\end{equation}

One would be puzzled that the $B_r$ is greater than $ B_{r}^{0}$. This is because we are assuming that all the flux is contained within the outer core
and the field energy in such a configuration is more than in the free field configuration. We note here that exact flux conservation is a simplifying assumption - we will comment on this later. The flux averaged energy locked in the screening field then follows,

\begin{equation}
E^{s}_{mag} = \frac{1}{8\pi} \int_{2 r_c}^{R}B_s(r)^2 \cdot 4 \pi r^2~dr
\end{equation}

Integrating this equation, substituting the time dependent expression for the radius (eq. \ref{r_time_dep}) and differentiating $ E^s_{mag}$ with respect to time, we get the rate of magnetic energy dissipation as

\begin{equation}
L_X = \frac{1}{16}~B_{c}^{2} r_{c}^{3}~\frac{\xi_2}{3.15 \times 10^{13}}~\frac{e^{-3 \xi_2 t} (1 - 4 e^{2 \xi_2 t} + 8 e^{3 \xi_2 t})^2}{(1 - 4 e^{2 \xi_2 t})^2}
\end{equation}

We had earlier fixed $t = 0$ in the ambipolar phase at $r = 2 r_c$.  On the other hand, in the surface field parameterization, for $t = 0$, $B(t) = B_0 (\sqrt{1  - e^{-x}}) = 0$ (where $x = \xi t$). Assuming that the time  $t$ for the surface field evolution is to match with the time as defined in the ambipolar evolution above,  we should have $r(t = 0) = 2 r_c$ and $B(t = 0) = 0$.\\

Consequently we note that as we move to $r = R$ in the ambipolar evolution, the surface field  $B(t)$ also evolves, and thus all the flux is not returned by the screening currents in the outer core ambipolar region -- contrary to the assumption of total flux conservation. The $B(t)$ evolution implies that the flux and field change in the crust and thus also at the surface. Strictly speaking. flux conservation  would suggest zero field outside $r$, so the surface field should remain zero until the screening boundary reaches the surface. As pointed out in the last paragraph this is not the case and thus our formulation is  aresonable approximation only when the flux contained within the screening radius is much larger than that penetrating the surfaceand can give rise to irregrity at the core crust boundary.\\

The decrease in electrical and thermal conductivity \citep{Gnedin} in going from the core to the crust will increase the temperature in the transition region thus increasing $\xi_2 = 1/\tau_B$, which has a very sensitive dependence on the temperature ($T^6 $ \cite{Passamonti,DipSoni}). We can estimate the temperature dependance of $\xi_2$ as we approach the crust at $R$ using the results of \cite{Kaminker}. The value of $\xi_2$ is determined at the data point when $\tau_{SD}$ and $B(t)$ are observed, which corresponds to a given value of $r_{obs}$. The results of \cite{Kaminker} suggest that the temperature change from the outer core to the inner crust is such that the ratio  $\Lambda(T) = T_{crust}/T_{outercore}$ is at least of order $10^{1/3}$. This suggests that the change in the value of $\xi_2$ from $r_{obs}$ to the radius of the inner crust $R$ is of order $100$.\\

By assuming a linear increase in $\xi_2$ as $r(t)$ changes, we carry out an integration in discrete steps of $r(t)$ from $r_{obs}$ to $R$, simultaneously changing the value of $\xi_2$ at each step. We use eq. \ref{t_r}   (for a given $\xi_2$ at each step) to calculate the time step. We then add up all the time steps till we get to $R$ to determine the  time $t_1$ for ambipolar diffusion to end at the crust. This graded increase in $\xi_2$ will augment the ambipolar dissipation rate -- it will first cause an increase in $L_X$ and then $L_X$ will decrease exponentially with time.\\

We find that the observed $ \tau_{SD}$ and $t$ from polar field evolution fall in the ambipolar regime, that is, $t < t_1$, where $t_1$ is the time when $r$ reaches the crust at $r = R$. We fix $\xi_2$ from $L_X\text{(observed)} = L_{ambipolar}$ in the ambipolar regime. At times larger than $ t > t_1$,  we switch to $L_{ohm}$ given in eq. \ref{L_ohm}. As expected, we get a discontinuity between $L_{ambipolar}$ at $t_1$ and $L_{ohm}$ at $t = t_1$, which we smoothen out.\\




However, we remind the reader this a heuristic calculation to make some approximate estimates of the ambipolar and ohmic contribution to $L_X$.\\

We shall present estimates for $L_X $ based solely on ambipolar diffusion for the two stars below

\begin{enumerate}
	\item 4U 0142+61, whose spin down age is large, $\tau_{SD} \sim 0.068$ Myr, small surface polar field, $B(t) \sim 1.34 \times 10^{14}$ G but large $L_X \text{(thermal)} \sim 2.5 \times 10^{34}$ erg s$^{-1}$ \citep{Rea2}.
	
	\item CXOU J010043.1-721134, a typical magnetar in the intermediate range, whose spin down age is $0.0068$ Myr, surface polar field, $B(t) \sim 3.9 \times 10^{14}$ G and $L_X \text{(thermal)} \sim 6.5 \times 10^{34}$ erg s$^{-1}$.
	
\end{enumerate}

Given these expressions for the instantaneous period and its derivative one can write down the dipolar spin down radiation energy loss rate for this model,

\begin{equation*}
\large \dot E = {\frac{4\pi^2 I \dot P}{P^3}}~\text{erg s$^{-1}$}
\end{equation*}

In this scenario the late time dipole radiation mediated spin down energy rate goes as $x^{-2}$, while $L_X$ falls as $ e^{-x} $. The figures indicate that, unlike the case of the dynamo model, in this model, $ \dot E/L_X $ starts being larger than $1$, then falls well below $1$ and finally becomes greater than $1$ as the spin down age goes up and period goes beyond $ P\sim 10 - 15$ s. This can be seen from the figures. The figures in the next subsection show how all the parameters of this model evolve with time.

\subsubsection*{Remarks \& Plots}

The screened core model is indeed rather different from the dynamo model in that two of the three stars, we need to consider ambipolar diffusion to provide the observed luminosity -- Ohmic dissipation in the crust is too small. For these stars, we have taken $L_X$ to come from ambipolar diffusion, upto the time $t_1$, when the screening currents have moved to the crust.\\


However, as in the dynamo model, for the remaining magnetar (SGR 1806-20), we have taken the full contribution of $L_X$ to come from Ohmic dissipation in the crust. Whereas the value of $\xi $ for the dynamo could be set by the observed $L_X$ to be rather small ($\xi \sim 13$ Myr$^{-1}$) compared to magnetic field scaled value ($\xi \sim 7500$ Myr$^{-1}$), in the screened model, we do not have such a relation, so we use the latter value of $\xi$.\\

In a later work, we intend to do a fuller treatment which will include the dissipation from neutrinos and simultaneously from ambipolar diffusion for SGR 1806-20, and the Ohmic dissipation for the magnetars 4U 0142+61, CXOU J010043.1-7211334. Our preliminary findings indicate that the late time and early time behaviour is similar to what we have found above.\\


\begin{center}
	\begin{table}[ht]
		\centering
		\caption{Screened Core Model Values}
		\label{my-label}
		\begin{tabular}{lllllll}
		\hline
		Magnetar & $x$ & $\xi$ (Myr$^{-1}$) & $\xi_2$ (Myr$^{-1}$) & $t/\tau$ & $t_1$ (kyr) & $B_0$ (G)\\ \hline
		4U0142+61 & 3.254 & 35.019 & 6.500 & 1.364 & 207.230 & $1.36 \times 10^{14}$\\
		CXOU J010043.1-721134 & 2.902 & 306.405 & 3.172 & 1.401 & 19.542 & $4.04 \times 10^{14}$\\
		SGR 1806-20 & 2.523 & 7203 & - & 1.447  & - & $2.04 \times 10^{15}$\\ \hline
		\end{tabular}
	\end{table}
\end{center}

\begin{figure}[H]
	\centering
	\begin{minipage}[b]{0.4\textwidth}
		\includegraphics[scale=0.06]{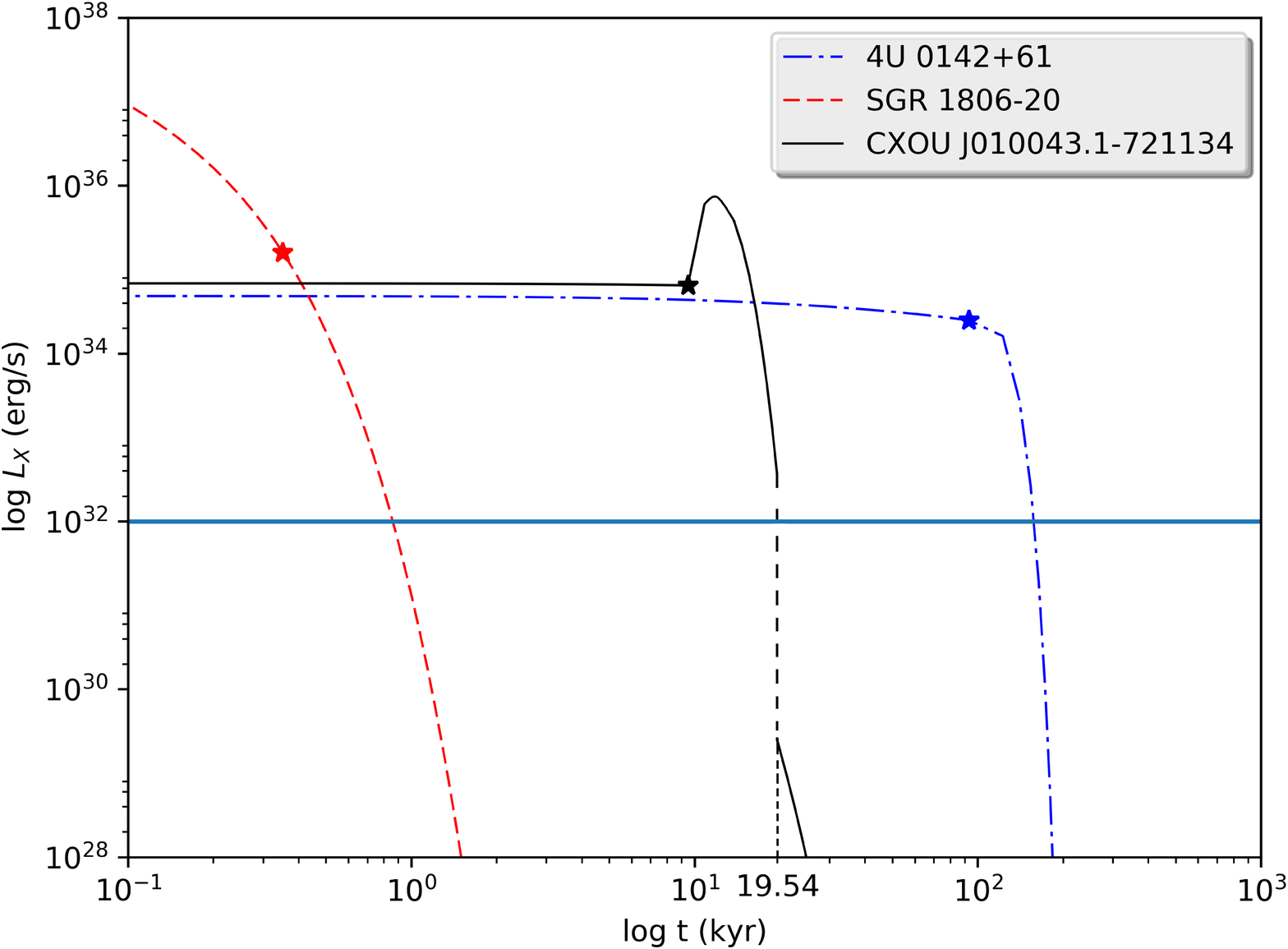}
		\caption{\footnotesize X-ray Luminosity Vs. Time}
	\end{minipage}
	\hfill
	\begin{minipage}[b]{0.4\textwidth}
		\includegraphics[scale=0.06]{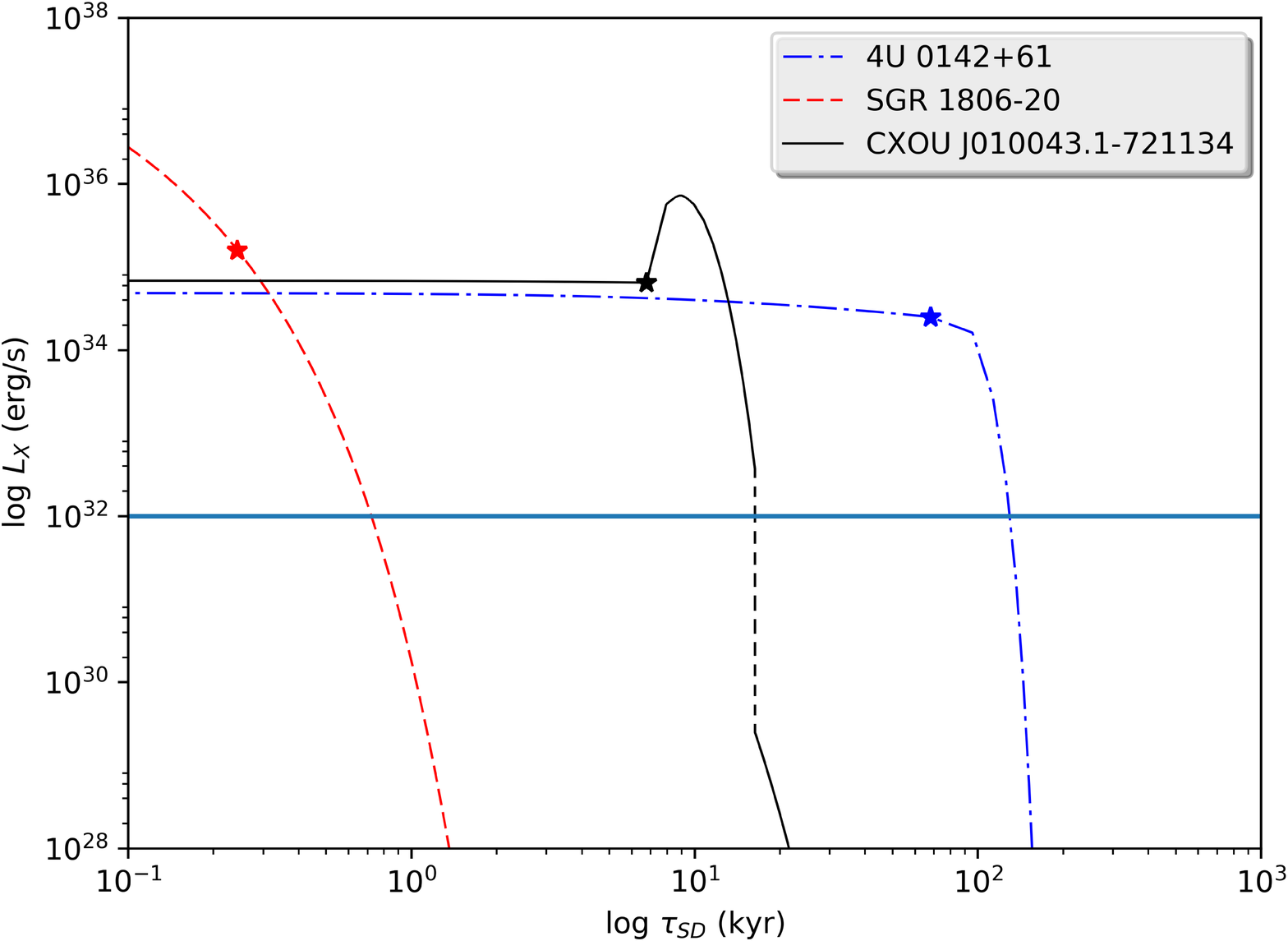}
		\caption{\footnotesize X-ray Luminosity Vs. Spin Down Age}
	\end{minipage}
\end{figure}

\begin{figure}[H]
	\centering
	\begin{minipage}[b]{0.4\textwidth}
		\includegraphics[scale=0.06]{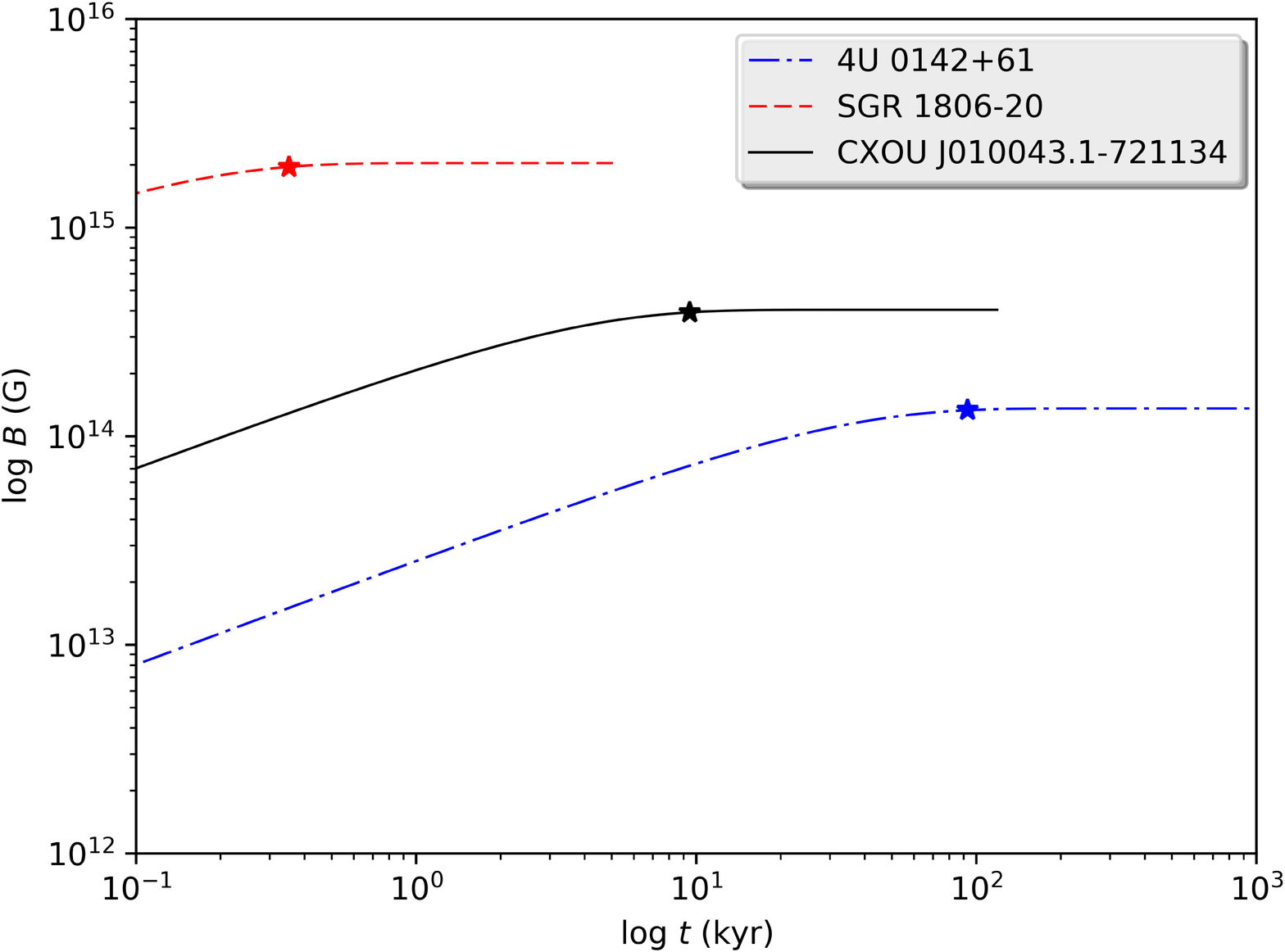}
		\caption{\footnotesize $B(t)$ Vs. Time}
		\vfill
	\end{minipage}
	\hfill
	\begin{minipage}[b]{0.4\textwidth}
		\includegraphics[scale=0.06]{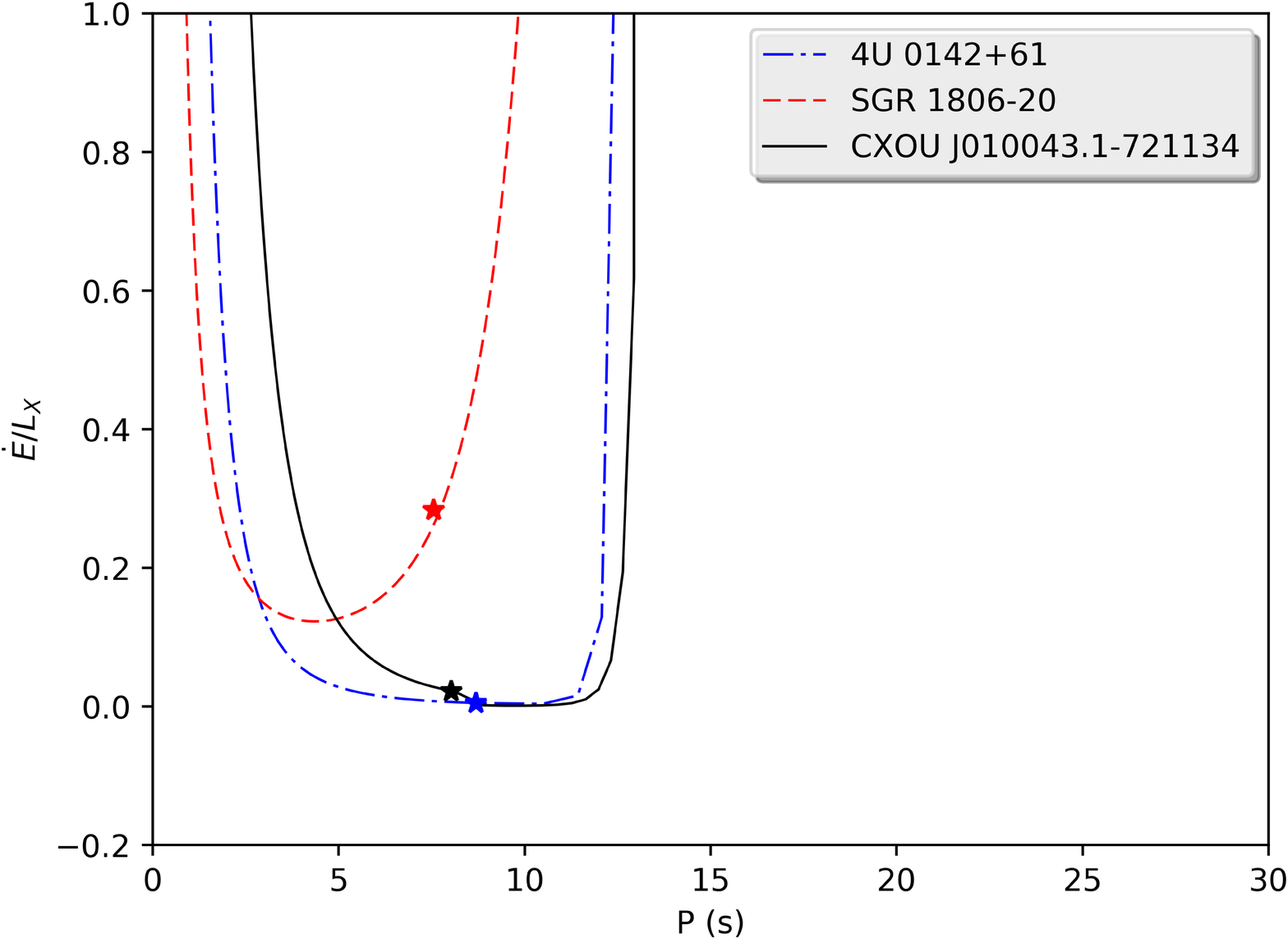}
		\caption{\footnotesize $\dot E/L_X$ Vs. Period}
	\end{minipage}
\end{figure}

\begin{figure}[H]
	\centering
	\begin{minipage}[b]{0.4\textwidth}
		\includegraphics[scale=0.06]{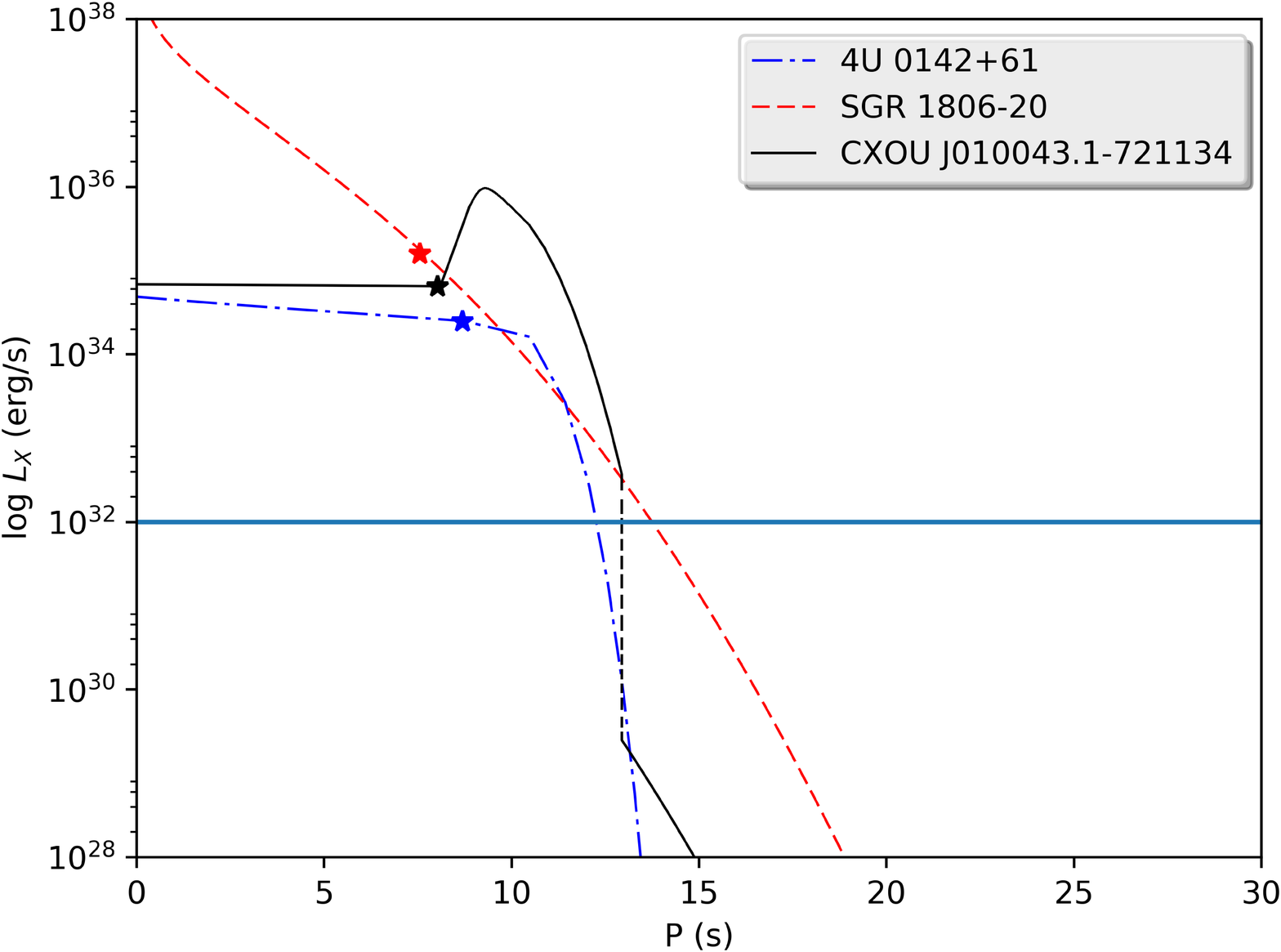}
		\caption{\footnotesize X-ray Luminosity Vs. Period.}
	\end{minipage}
	\hfill
	\begin{minipage}[b]{0.4\textwidth}
		\includegraphics[scale=0.06]{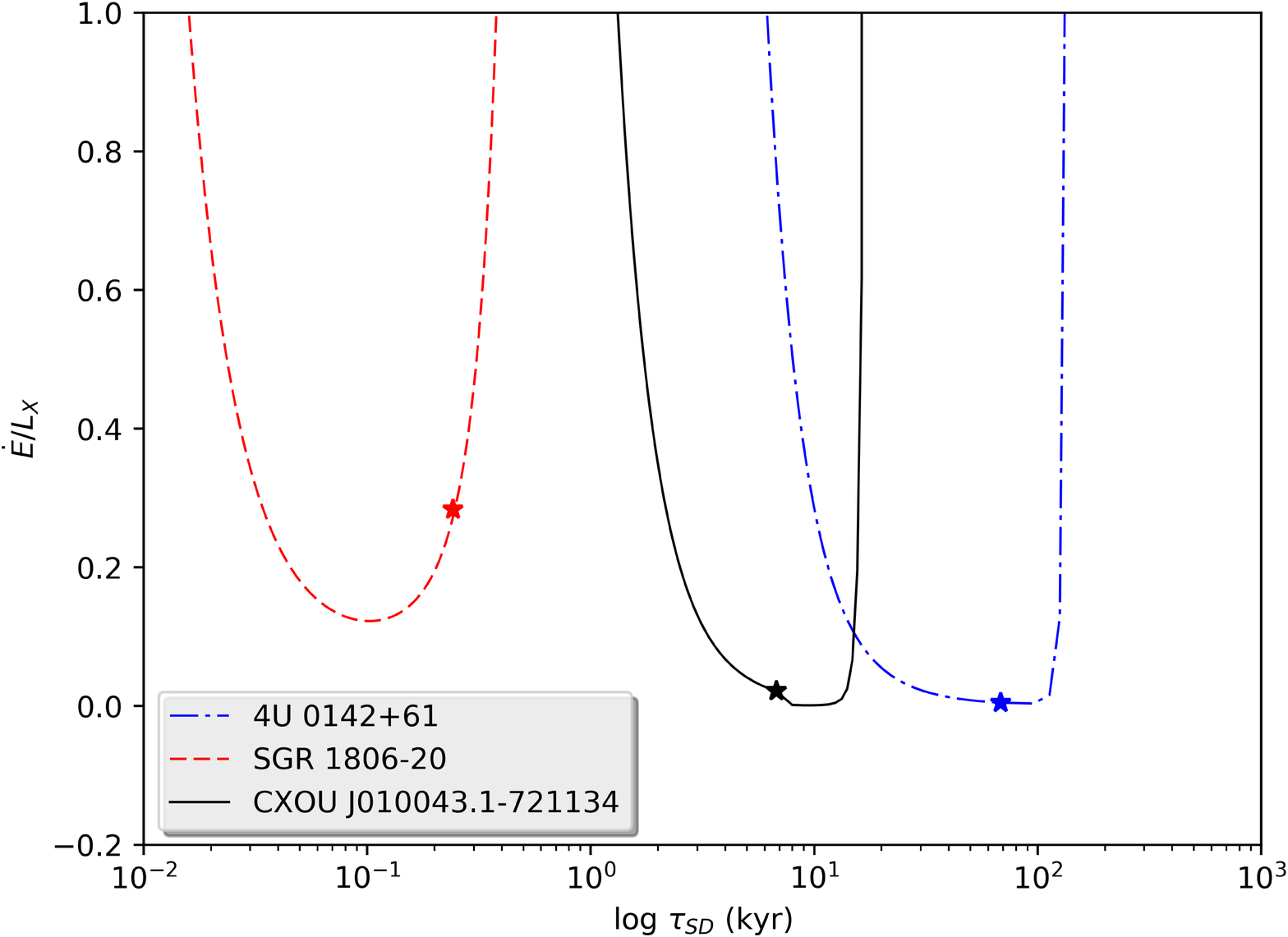}
		\caption{\footnotesize $\dot E/L_X$ Vs. Spin Down Age}
	\end{minipage}
\end{figure}

The horizontal line in the plots with X-ray luminosity corresponds to a value of $L_X = 10^{32}$ erg s$^{-1}$. This bound is marked due to magnetars becoming potentially difficult to observe under this value. One should also note that the black dashed line in the curve for CXOU J010043.1-721134 shows the discontinuity in going from $L_{ambipolar}$ to $L_{ohm}$. In the case of 4U 0142+61, the line of discontinuity is not seen since at $t_1 = 207.23$ kyr, the luminosity drops below $10^{28}$ erg s$^{-1}$ when the transition from $L_{ambipolar}$ to $L_{ohm}$ occurs.

\section{Evidence for the absence of magnetar phenomena for large period ($P > 12$ s)}
\label{sec6}

We emphasize one of the important differences between the dynamo model and this one. Both models start with $L_X < \dot E$, followed by the reversal $L_X > \dot E$. However, for the screened core model, $L_X$ drops faster and so for $ P > 10-20$ s, we get another reversal $L_X < \dot E$ for late times whereas in the dynamo model, $L_X > \dot E$ for late times. \textit{For us, this is the characteristic signature of the magnetar phenomenon.} If we plot $\dot E/L_X$ versus the period of the star, $P$, we find a \textbf{U curve} that dips below $1$ for a small interval in the period. This is interval in which magnetars live.\\

In the dynamo model where the X-ray radiation comes from the gradual dissipation of the currents that support the magnetic field, we may not expect a cessation or sharp decline of the X-ray radiation. Actually, the late time behaviour $L_X > \dot E$ with $L_X > 10^{33}$ erg s$^{-1}$ indicated for the dynamo model is not seen. It has been pointed out by \cite{Rea1} for accreting binary X-ray pulsars with smaller magnetic fields that it is common to find periods even as large as a few minutes. However, in this case, the X-ray emission comes from a different source -- from heating of the surface from the matter accreting from one star on to the other. This emission continues even though the periods are very large.\\

In the screened core model, the dissipation of screening currents in the core and crust are assumed to be the primary source of quasi steady X-ray emission. One of the significant features of the screened core model is that the screening currents get rapidly dissipated as $B_{polar}$ reaches a maximum. After this, there is no further quasi steady X-ray radiation. For larger periods, we see a turn around as $L_X < \dot E$ with a sharp decline in $L_X$ in the screened core model. We interpret the non observation of magnetars with periods larger than $12$ s as meaning that there is a near cessation of X-ray emissions for larger periods (not that larger periods do not exist). Also, the observed data shows that magnetars with large periods continue to have high magnetic fields. This would support our model's predictions.\\

One question that remains for the screened core model is that after the screening currents are dissipated and X-ray radiation wanes, we are left with high (highest) polar magnetic fields and large periods. We may expect that magnetars with periods $P > 12 - 20$ s should be visible for such high $B_{polar}$ in the radio spectrum. The end state of magnetars in the screened model leaves behind a large surface polar field ($B_0 \sim 10^{14-15}$ G) but also a large period ($P> 10 - 20$ s), but since they fall around the so called death line, they are not seen. However, we are hopeful that for the highest fields seen ($B_0 > 10^{14-15}$ G), we could potentially escape the death line to permit observations.

\section{Discussion}
We have made an effort to look for features and signatures of magnetars in two different models for a very complex magnetohydrodynamic system. We have worked with many simplifying assumptions:

\begin{enumerate}

\item $B_{polar}$ is parametrized in the two models by exponential decay of the polar magnetic field in the dynamo model and an exponential decay of screening currents in the screened core model.

\item We have assumed for simplicity that both ambipolar ($\xi_2$) and Ohmic ($\xi$) dissipation timescales are constant parameters (except for the sensitive temperature dependence of the ambipolar timescale in going from the outer core to the inner crust), that are set by field dependent conductivities in the crust and the observed parameters of the magnetar in question.

\item We have also assumed that the core fields (but not the core radii) are constant for all magnetars.

\end{enumerate}

Our results are therefore to be viewed as providing features of the magnetars evolution and not reliable quantitative estimates.\\

We have found that the Hall drift model falls short of observed $L_X$ especially when compared to canonical magnetars. We also found that the resistive layer model may not provide periods of $P < 12$ s observed for high field magnetars with $B_{polar} > 10^{15}$ G. It also does not show the sharp fall off in $L_X$, which is observed for $P > 15$ s. Actually, this model is very similar to the dynamo model (below).\\

In the dynamo model, we do get period saturation for constant $\xi$, but not for a magnetic field dependent $\xi$. However, even for constant $\xi$, as stated before, for large fields ($ B > 10^{15}$ G), the period is likely to saturate at values much larger than observed for magnetars. Also, whereas, observations indicate that $L_X$ dies out fast for magnetars after periods of $P \gtrsim 12-20$ s, that does not seem to the case for the dynamo model. Actually, $L_X > \dot E$ for large spin down age (and period) which is not seen for magnetars.\\

In our screened core model, there is no cap on the period which goes as $x^{1/2}$ at late times. It seems that this is in contradiction with the observation of the cap on magnetar periods. We, however, must remember that observations indicate a near cessation of X-ray emission from magnetars when periods exceed $\sim 12-20$ s -- not that they do not have periods larger than $12$ s. We read the observations as such: magnetars show a cessation of  X-ray emissions approximately after periods of $12-20$ s regardless of their polar magnetic fields -- actually magnetars do not show any decline in the polar magnetic fields even at the largest periods. But since they are radio quiet, their subsequent period evolution cannot be tracked.\\

A notable difference between the two models is the plot of $\dot E/L_X$ versus $P$ or $\tau_{SD}$ where we found a \textbf{U curve} that dips below $1$ for an interval in $P$ or $\tau_{SD}$ for the screened core model but not for the dynamo model. This is the interval in which magnetars are found.\\

These considerations indicate that the screened core model, which posits the existence of a dynamically magnetised inner core, offers a plausible explanation for magnetar phenomena. This would have implications for the equation of state of neutron stars.

\bibliography{references}
\bibliographystyle{apj}

\end{document}